\documentclass[fleqn,10pt]{wlscirep}
\usepackage[utf8]{inputenc}
\usepackage[T1]{fontenc}
\usepackage{xcolor} 
\usepackage[normalem]{ulem}
\usepackage{tikz}
\usepackage[capitalise]{cleveref}
\usepackage{braket}

\newcommand\TikCircle[1][2.5]{\tikz[baseline=-#1]{\draw[thick](0,0)circle[radius=#1mm];
\draw[thick](0,0)circle[radius=2mm];}}
\newcommand\TikCircleTwo[1][2.5]{\tikz[baseline=-#1]{\draw[thick](0,0)circle[radius=#1mm];
\draw[thick](0,0)circle[radius=2mm];
\node[rotate=45, thick] at (0.05,0.05) {\uwave{\hspace{0.4cm}}};}}

\newcommand\TikPair[1][2.5]{\tikz[baseline=-#1]{
\draw[thick](0,0)--(0.3,0.25);
\draw[thick](0.075,0)--(0.3,0.175);
\draw[thick](0,0)--(0.3,-0.25);
\draw[thick](0.075,0)--(0.3,-0.175);}}

\newcommand{\be}{\begin{equation}}
\newcommand{\ee}{\end{equation}}
\newcommand{\bi}{\begin{itemize}}
\newcommand{\ei}{\end{itemize}}
\newcommand{\vphi}{\varphi}	
\newcommand{\bea}{\begin{eqnarray}}
\newcommand{\eea}{\end{eqnarray}}
\newcommand{\eps}{\varepsilon}
\newcommand{\vkap}{\varkappa}

\newcommand{\tsf}[1]{\textsf{#1}}
\newcommand{\trm}[1]{\textrm{#1}}

\newcommand{\prob}{\tsf{P}}
\newcommand{\ud}{\mathrm{d}}

\newcommand{\RE}{{\textrm{Re}\,}}
\newcommand{\IM}{{\textrm{Im}\,}}

\newcommand{\figref}[1]{Fig. \ref{#1}}
\newcommand{\figrefa}[1]{Fig. \ref{#1}a}
\newcommand{\figrefb}[1]{Fig. \ref{#1}b}

\newcommand{\eqnref}[1]{Eq. (\ref{#1})}

\newcommand{\lcperp}{{\scriptscriptstyle \perp}}
\newcommand{\sech}{\text{sech}}

\title{Strong-field vacuum polarisation with high energy lasers}

\author[1,*]{A.~J.~Macleod}
\author[2]{J.~P.~Edwards}
\author[2]{T.~Heinzl}
\author[2,3]{B.~King}
\author[1,4]{S.~V.~Bulanov}

\affil[1]{ELI Beamlines Facility, The Extreme Light Infrastructure ERIC, Doln\'{i} B\v{r}e\v{z}any, Czech Republic}
\affil[2]{Centre for Mathematical Sciences, University of Plymouth, Plymouth, PL4 8AA, United
Kingdom}
\affil[3]{Deutsches Elektronen-Synchrotron DESY, Notkestr. 85, 22607 Hamburg, Germany}
\affil[4]{National Institutes for Quantum and Radiological Science and Technology (QST), Kansai Photon Science Institute, 8-1-7 Umemidai, Kizugawa, Kyoto 619-0215, Japan}

\affil[*]{alexander.macleod@eli-beams.eu}

\begin{abstract}
When photons propagate in vacuum they may fluctuate into matter pairs thus allowing the vacuum to be polarised.
This \emph{linear} effect leads to charge screening and renormalisation. 
When exposed to an intense background field a \emph{nonlinear} effect can arise when the vacuum is polarised by higher powers of the background.
This nonlinearity breaks the superposition principle of classical electrodynamics, allowing for light-by-light scattering of probe and background photons mediated through virtual pairs dressed by the background. Vacuum polarisation is a \emph{strong-field} effect when all orders of interaction between the virtual pair and the background must be taken into account. 
In this investigation we show that multiple scattering processes of this type may be observed by utilising high-energy laser pulses with long pulse duration, such as are available at facilities like ELI Beamlines. 
In combination with appropriate sources of high-energy probe photons, multiple probe-background light-by-light scattering allows for testing the genuine nonlinear regime of strong-field QED. This provides access to the uncharted nonperturbative regime beyond the weak-field limit.
\end{abstract}
\begin{document}

\flushbottom
\maketitle
\thispagestyle{empty}

\section{Introduction \label{sec:Intro}}

A well-known principle of classical electrodynamics is that electromagnetic fields can be linearly superposed: they do not interact. 
This is true in vacuum and in linear media where the dielectric response is field independent. 
In strong fields the linear approximation becomes insufficient, requiring the inclusion of higher orders in field magnitude. 
As a result, permittivity or permeability become field dependent, and one enters the field of nonlinear optics where the superposition principle no longer holds (see e.g.\ the lucid text \cite{Roemer:2005}). 
An analogous situation arises in quantum electrodynamics (QED) which couples photons to charged matter particles. Hence,
photons propagating in vacuum may fluctuate into matter / anti-matter pairs, making the vacuum polarisable by an electromagnetic background.
At lowest order in the coupling, given by the fine structure constant $\alpha = e^2/4\pi = 1/137$, this is a linear effect leading to charge screening and hence a modification of the Coulomb interaction, for instance. 
This has been known since the early days of QED from the calculations of Dirac \cite{Dirac:1934a}, Heisenberg \cite{Heisenberg:1934pza}, Uehling \cite{Uehling:1935uj} and Serber \cite{Serber:1935ui}, as well as Pauli and Rose \cite{Pauli:1936zz}. 
(Serber even entitled his paper `Linear modifications in the Maxwell field equations'.)

Going to higher order in $\alpha$, QED becomes a nonlinear theory. This is most easily seen by integrating out the fermions which results in a  quantum effective action of self-interacting photons, described by higher powers of the electromagnetic field strength. This was first studied by Heisenberg and his students \cite{Heisenberg:1936nmg,Euler:1935qgl,Euler:1935zz}, focussing on the low-energy limit.  The coupling to matter proceeds entirely through virtual channels, namely through vacuum polarisation loops. 
A transparent interpretation was given by Weisskopf \cite{Weisskopf:1936hya} who stated explicitly that below the pair threshold ``\emph{the electromagnetic properties of the vacuum can be represented by a field dependent electric and magnetic polarisability of empty space which leads, for example, to the refraction of light in electric fields or to the scattering of light from light}''.  

In modern terms, the situation can be described by a low-energy effective field theory of self-interacting photons, whose Lagrangian can schematically be written as a double expansion in derivatives and field strengths,
\be
  \mathcal{L}_\mathrm{eff} = \sum_{\sigma, n} c_{\sigma,n} \,\, 
  \partial^{(\sigma)} \mathcal{O}^n \; .
\ee
The coefficients $c_{\sigma,n}$ are low-energy constants with indices $\sigma$ and $n$ counting the number of derivatives and field strengths in the operator $\mathcal{O}$, respectively.  
More explicitly, with all Lorentz indices appropriately contracted, one finds (after absorbing a logarithmic term via charge renormalisation),
\be \label{L.EFF}
  \mathcal{L}_\mathrm{eff} = c_{02} F_{\mu\nu}F^{\mu\nu} + 
  c_{22} F_{\mu\nu} \Box F^{\mu\nu} + c_{04,1} 
   (F_{\mu\nu}F^{\mu\nu})^2  + c_{04,2} 
   (F_{\mu\nu}\tilde{F}^{\mu\nu})^2 + \sum_{i=1}^7 c_{24,i} L_i + 
   \ldots \; .
\ee
The first term corresponds to Maxwell theory with field strength $F_{\mu\nu}$ (and dual $\tilde{F}_{\mu\nu}$); the second term is the correction first calculated by Dirac and Heisenberg \cite{Dirac:1934a,Heisenberg:1934pza}. 
Both are quadratic in field strength ($n=2$), implying a first-order equation of motion. 
Hence, the corresponding electrodynamical theories are linear\footnote{Historically, the theory with the first derivative correction has been suggested as a `fundamental' theory by Bopp \cite{Bopp:1940} and Podolsky \cite{Podolsky:1942zz} and now bears their name.}. 
The \emph{nonlinear} corrections have at least $n = 4$.
The terms without derivatives ($\sigma = 0$) correspond to the leading order Heisenberg-Euler Lagrangian \cite{Heisenberg:1936nmg}, while the $L_i$ terms each contain two derivatives and contractions of four field strength tensors\cite{Gusynin:1998bt}.
Power counting and dimensional analysis imply that the nonlinear coefficients $c_{04}$ are of order $\alpha^2/m^4 = \alpha /4\pi E_S^2$, thus inversely proportional to the square of the Sauter-Schwinger field, $E_S \equiv m^2/e \simeq 1.3 \times 10^{18}$ V/m which defines the critical field magnitude for pair creation\cite{Sauter:1931zz,Schwinger:1951nm}. 
This suggests that nonlinear effects involving four photons will only be measurable in the strong-field domain when the ambient fields approach the Sauter-Schwinger limit.  

In this paper, we will go beyond both the weak-field and low-energy limits by assuming (i) a  \emph{strong background} field and (ii) to probe the latter with \emph{high-energy} photons. 
For this purpose, the derivative expansion of the effective Lagrangian (\ref{L.EFF}) is insufficient and must be resummed. 
To do so (at least partially), we write the field strength in terms of a gauge potential $\mathsf{A}$ such that $F_{\mu\nu} = \partial_\mu \mathsf{A}_\nu - \partial_\nu \mathsf{A}_\mu$, and split it into (a dominant) background, $\mathcal{A}$ and a small perturbation, $A$, i.e.\ $\mathsf{A}_\mu = \mathcal{A}_{\mu} + A_\mu$.
The effective Lagrangian now takes the form of a Taylor expansion in the fluctuation $A$,
\bea \label{POL.TENSOR}
  \mathcal{L}_\mathrm{eff}[\mathcal{A}, A] = \mathcal{L}_\mathrm{eff}[\mathcal{A}, 0] + A_\mu J^\mu [\mathcal{A}] + \frac{1}{2} A_\mu \left(\Box g^{\mu\nu} + \Pi^{\mu\nu}[\mathcal{A}] \right) A_\nu + \ldots \; , 
\eea
with vacuum current $J^\mu$ and polarisation tensor $\Pi^{\mu\nu}$ both depending on the strong background field $\mathcal{A}$.
The polarisation tensor encodes the electromagnetic response of the background to an electromagnetic probe, in particular polarisation effects, and is thus aptly named. 
In the limits of interest, large probe energy and background field strength, we need to know the polarisation tensor to all orders in probe momentum, $k$, and background field $\mathcal{A}$. We note in passing that the low-energy effective Lagrangian (\ref{L.EFF}) yields a polarisation tensor of the form
\be
   \Pi^{\mu\nu}[\mathcal{A}] = c_{22} \, k^2 k^\mu k^\nu + c_{04,1} \, (\mathcal{F}k)^\mu (\mathcal{F} k)^\nu + c_{04,2} \, (\tilde{\mathcal{F}}k)^\mu (\tilde{\mathcal{F}} k)^\nu + \ldots \; , 
   \qquad \mathcal{F} \equiv d\mathcal{A} \; ,  
\ee
which clearly exhibits the first few terms of the double expansion just mentioned.
For a constant crossed field background (electric and magnetic fields orthogonal and of the same magnitude), the polarisation tensor has first been calculated by Toll\cite{Toll:1952rq}. He then determined the vacuum refractive indices (in terms of the eigenvalues of the polarisation tensor) and pointed out the relevance of the optical theorem in accounting for photon absorption (i.e.\ pair creation). The latter is described by the imaginary parts of the refractive indices, of which there are two, one parallel and one orthogonal to the preferred direction of the intense background field (`vacuum birefringence'). 

The advent of high-power lasers with ever increasing intensities has kindled renewed interest in nonlinear phenomena associated with photons scattering off electromagnetic backgrounds\cite{Lundstrom:2005za,Heinzl:2006xc,DiPiazza:2006pr}. 
It is impossible to cite all individual contributions in this context, so we refer the reader to recent review articles\cite{King:2015tba,Karbstein:2019oej,Fedotov:2022ely} for a more complete bibliography, and to\cite{Fedotov:2022ely,Ritus.J.Sov.Laser.Res.1985,Marklund.RevModPhys.78.591.2006,Ehlotzky.Rep.Prog.Phys.72.046401.2009,DiPiazza.RevModPhys.84.1177.2012,Narozhny.Cont.Phys.56.3.2015,Zhang.PoP.27.5.050601.2020,Gonoskov.RevModPhys.94.045001.2022} for more general reviews of strong-field quantum electrodynamics.
From a quantum field theory perspective, the leading (i.e.\ weak-field) contribution to the polarisation tensor in (\ref{POL.TENSOR}) is a light-by-light scattering diagram corresponding to the process ``probe + background $\to$ probe$^{\prime}$ + background''.
The study of this process has a long history \cite{Euler:1935zz, Euler:1935qgl, Akhiezer:1936vzu, Achieser:1937ywd, Karplus:1950zza,DeTollis:1965vna, Cheng:1970ef}, which is (at least in part) covered in the overview articles\cite{Scharnhorst:1998ix,Scharnhorst:2017wzh,Ahmadiniaz:2020jgo}.
Toll's dispersive and absorptive processes in probe-background interactions can be interpreted as elastic and inelastic light-by-light scattering events. 
Vacuum birefringence then corresponds to elastic scattering where probe photons flip their helicity when scattering off intense linearly polarised electromagnetic waves \cite{Dinu:2013gaa,Dinu:2014tsa,Bragin:2017yau,Aleksandrov:2023toy}. 
Pair creation, the transformation of probe photons into real electron-positron pairs, represents an inelastic process which, however, requires sufficiently intense fields and/or centre of mass energies. 
The inelastic process has so far only been observed experimentally in the linear weak-field regime by the E144 experiment \cite{Burke:1997ew}, but there are planned experiments, E320 (at SLAC) and LUXE\cite{Abramowicz:2021zja} (at DESY), to measure this for the first time in the nonperturbative strong-field regime. 

When the electromagnetic background is given by a Coulomb field, the corresponding background photons are virtual, carrying off-shell momenta $q$ with $q^2 = - \mathbf{q}^2 < 0$. 
The associated scattering process (Delbr\"{u}ck scattering\cite{Delbruck:1933pla}) was first measured decades ago\cite{Schumacher:1975kv}. 
More recently, light-by-light scattering in boosted Coulomb fields (with quasi-real photons) has been observed at higher energies in the ATLAS \cite{ATLAS:2017fur,ATLAS:2019azn} and CMS \cite{CMS:2018erd} experiments. 
It should also be emphasised that virtual light-by-light scattering diagrams contribute at third order to lepton g-factors\cite{Laporta:1991zw}. 
For the electron, they amount to about 30\% of the total contribution at this order\cite{Schlenvoigt:2016jrd}. 
The notorious hadronic light-by-light scattering diagrams are still not fully under control, in particular for the muon, and huge efforts continue to be invested into their determination\cite{Aoyama:2020ynm}.

Vacuum birefringence due to the scattering of real \emph{on-shell} photons, however, has yet to be measured, although evidence has appeared for its effect in an astrophysical context \cite{Mignani:2016fwz}. 
So far, the most sensitive lab measurement is from the PVLAS experiment\cite{Ejlli:2020yhk, Bakalov:1998hnn}, which reached a sensitivity for the refractive index measurement of just a factor of $50$ below the QED prediction. 
Much work has been published on the subject of colliding intense optical pulses with one another and also scenarios involving an xray free electron laser (XFEL) (see e.g. reviews \cite{King:2015tba,Karbstein:2019oej,Fedotov:2022ely}), which can be described using the leading non-derivative correction in the low-energy expansion of the effective action, \cref{{L.EFF}}. Indeed there are planned experiments such as HIBEF \cite{Schlenvoigt:2016jrd} (possibly using a Coulomb-assisted process \cite{Ahmadiniaz:2020kpl}) and the station of extreme light at the Shanghai Coherent Light Facility\cite{Shen:2018lbq} to measure vacuum polarisation  scattering of real photons at these energies.

In this work, we are interested in a different regime, in which the centre-of-mass energy is close to the pair rest energy, where all orders of interaction between the background and the polarised virtual pair must be taken into account. This parameter regime has been investigated theoretically \cite{Dinu:2013gaa,Dinu:2014tsa}, employing scenarios that combine a high-energy gamma source with an intense optical laser to measure vacuum polarisation effects \cite{Meuren:2014kla,King:2016jnl,Bragin:2017yau,Borysov:2022cwc}.
Here, we analyse the feasibility of scattering high-energy photons off intense laser pulses to observe not only helicity flip (and hence vacuum birefringence) but to allow measurement of \emph{multiple} photon scattering, and to do so in a regime where the coupling of the photon to the background is nonperturbative in the charge-field interaction. 
This is achieved by using a semi-classical polarisation-operator approach which describes strong-field effects on the evolution of the polarisation dependent phase in the background field. 
Multiple scattering events can be expected when the mean free path of the probe photon is small compared to the laser pulse length. Introducing the peak dimensionless intensity parameter of the electromagnetic field,  $\xi_{0}$, the peak quantum nonlinearity parameter of the photon, $\chi_{0}$, and denoting by $\Phi$ the laser pulse duration, this is when when the parameter
\begin{equation} \label{eqn:Delta.el}
    \Delta^{\tiny\tsf{el.}} = \alpha\xi_{0}\chi_{0}\Phi/30\pi \gtrsim 1\,.
\end{equation}
This suggests three possible routes to observe multiple scattering: (i) increasing the intensity parameter, (ii) increasing the photon energy, $\omega_{\gamma}$ (since $\chi_{0} \propto \omega_{\gamma}$), or (iii) increasing the pulse duration.
However, as $\xi_{0}$ and/or $\chi_{0}$ are increased, inelastic scattering, characterised by 
\begin{equation} \label{eqn:Delta.inel}
    \Delta^{\tiny\tsf{inel.}} = \alpha\xi_{0}\Phi\exp(-8/3\chi_{0}) 
\end{equation} 
becomes important ($\Delta^{\tiny\tsf{inel.}} \gtrsim$ 1) and the probability that the photon instead decays into an electron-positron pair increases (short photon `life-time')\footnote{
Significant pair production would be detrimental because (a) initial photons are removed, reducing the measurable signal and (b) the approach used here would not account for secondary photon production from pairs accelerated by the laser. Here the consistency of our approach is examined through the optical theorem relating the imaginary part of the vacuum polarisation amplitude to nonlinear Breit-Wheeler pair creation. 
We use this to identify where pair creation becomes significant through the differing signals predicted with and without pair creation, obtaining constraints on the experimental parameters and a quantitative method for assessing the validity of our approximations.}.
To observe multiple photon scattering in an intense laser pulse, one should therefore keep $\chi_{0} < 1$ to ensure the probability of pair creation is exponentially suppressed, such that the most effective route towards the ideal scenario, $\Delta^{\tiny\tsf{el.}} \gtrsim 1$ and $\Delta^{\tiny\tsf{inel.}} \ll 1$, is to use long pulse duration, $\Phi$.
This can be realised by exploiting a unique feature of the upcoming L4-ATON laser at the ELI Beamlines facility~\cite{Weber.Mat.Rad.Ex.2017}, which will offer pulses with large energies of 1.5~kJ.
(While the primary goal of this laser system is to achieve peak powers of 10~PW by pulse compression to a duration of 150~fs, long pulse duration remains another option.)
The minimum pulse duration that can be achieved by a high-power laser is determined by the spectral bandwidth of the amplification medium, but in principle there is no fundamental constraint on the \emph{maximum} duration which can be achieved.
The large available pulse energy of the L4-ATON laser at ELI Beamlines will allow for exceptionally long pulse durations of the order of $\text{ps}-\text{ns}$~\cite{Jourdain.Mat.Rad.Ex.2021}, while still maintaining sufficiently large $\xi_{0}$ for strong-field effects to be induced.

The paper is outlined as follows.
In \cref{sec:Index} we discuss our model of photon polarisation, reviewing the polarisation operator and its relationship to the accumulated phase of photons propagating in strong electromagnetic fields.
We work within the locally constant field approximation in the long-pulse limit.
We also outline the key physical observable, the Stokes parameter, which gives the asymmetry of photons measured with polarisations in two different basis modes.
In \cref{sec:Multiple} we describe how the all-orders Stokes parameter can be perturbatively expanded to give a length-$n$ chain of order $\alpha$ photon scattering events, and discuss how this can be used to indicate that a photon has undergone $(2n+1)$-fold scattering.
We outline an experimental scenario for measuring vacuum birefringence through the measurement of the Stokes parameter in \cref{sec:Exp}.
The experimental setup consists of a photon production stage, which we envisage will use the (inverse) nonlinear Compton scattering of high-energy electrons on a weak laser pulse to produce highly polarised photons, followed by a birefringence stage, where the photons interact with a second \emph{high-energy} laser pulse stretched to picosecond durations.
This scenario is motivated largely by the upcoming capabilities of the ELI Beamlines L4-ATON laser system.
In \cref{sec:Monoenergetic} we consider the idealised case of monoenergetic photons interacting with a linearly polarised plane-wave pulse.
This gives us an opportunity to demonstrate the key physical observables and the signatures of multiple scattering which our suggested experimental set-up could explore.
These results are then generalised to the case of non-monoenergetic photons scattering on plane-wave pulses in \cref{sec:EnergySpectra}.
We consider two different cases of photon energy distribution: idealised Gaussian distributions in \cref{sec:EnergySpectraGaussian} and nonlinear Compton scattering photon distributions in \cref{sec:EnergySpectraNLC}.
This allows us to demonstrate how the physical observables are changed by the presence of a photon energy spectrum and the optimal conditions for ensuring a robust experimental signal.
In \cref{sec:Focus} we include the effect of focussed laser pulses on the key observables in the interaction of nonlinear Compton scattered photons.
Finally in \cref{sec:Discussion} we summarise and discuss our results, and highlight future work which could be explored.

\section{Refractive Index Approach} \label{sec:Index}

For the purposes of this paper, we focus on the case of a probe photon scattering in a plane-wave background, $\mathcal{A}=\mathcal{A}(\vphi)$ with phase $\vphi=\vkap\cdot x$ and wavevector $\vkap$ obeying $\vkap\cdot\vkap = 0$ and $\vkap\cdot \mathcal{A}=0$.
Given this type of background, the current $J[\mathcal{A}]$, represented by a tadpole diagram, becomes \emph{linear} in the background field, $\mathcal{A}$\cite{Ahmadiniaz:2019nhk,DiPiazza:2022lij}. As a result, the current term, $A \cdot J$, in the effective Lagrangian can be absorbed by charge renormalisation, and should be, as there is no gauge invariant scalar of this form for plane wave (null) fields.
Varying the effective Lagrangian $\mathcal{L}_{\mathrm{eff}}$ from \eqnref{POL.TENSOR} with respect to the probe field $A$, one thus arrives at the modified wave equation (in Lorenz gauge) for the probe\footnote{Similarly to the current term, the vacuum polarisation is also renormalised by subtracting the zero-field limit $\Pi^{\mu\nu}[\mathcal{A}] \rightarrow \Pi^{\mu\nu}[\mathcal{A}] -\Pi^{\mu\nu}[0]$, which is left implicit throughout. Its resulting low energy limit being finite is important for the removal of the current term -- see \cite{DiPiazza:2022lij}.}:
\be
\square\, A^{\mu} = \Pi^{\mu\nu}[\mathcal{A}]A_{\nu}.
\label{eqn:we3}
\ee
Suppose the classical phase of the photon in vacuum is $\phi_{0} = k_{0} \cdot x$, where $k_{0}^{2}=0$. Then \eqnref{eqn:we3} can be solved with the ansatz:
\bea \label{eqn:Adef}
A^{\mu}_{j} = C\eps^{\mu}_{j}f_{j}(\vphi)\mbox{e}^{i \phi_{0}}, \label{eqn:A1a}
\eea 
where $C$ is a dimensional constant. One eventually finds: 
\be 
    f_{j}(\vphi) 
    = 
    \mbox{e}^{i\delta\phi_{j}(\vphi)} 
    \quad \trm{where} \quad
    \delta\phi_{j}(\vphi) 
    = 
    \frac{1}{\eta}\int^{\vphi} \frac{\eps_{j}^{\ast}\cdot \Pi[\mathcal{A}(y)] \cdot \eps_{j}}{2m^{2}}dy, 
    \label{eqn:deldef1}
\ee
where $\eta=\vkap\cdot k/m^{2}$ is the energy parameter of the photon and $\delta\phi_{j}(\varphi)$ is a field-induced phase lag for the polarisation mode $\varepsilon_{j}$. 
In the case of a constant crossed field background, the integral in \eqnref{eqn:deldef1} becomes trivial, the $\vkap$-dependency cancels, and the exponential terms in \eqnref{eqn:A1a} can be collected and the solution written as:
\bea \label{eqn:Adef2}
A^{\mu}_{j} = C\eps^{\mu}_{j}\mbox{e}^{i k_{j}\cdot x},\quad \trm{where} \quad k_{j}^{2} = \eps_{j}^{\ast}\cdot \Pi \cdot \eps_{j} \equiv \left(k_{j}^{0}\right)^{2}(1-n_{j}^{2}),  \label{eqn:A1a2}
\eea
with $n_{j}$ the \emph{vacuum refractive index} of photon polarisation state $\eps_{j}$. 
Writing the change in vacuum refractive index as $\delta n_{j} = n_{j}-1$ and demanding $\delta n_{j} \rightarrow 0$ as $k^{2} \rightarrow 0$ the phase lag, $\delta\phi_{j}(\vphi) = k_{j}\cdot x - k_{0}\cdot x$,  of the photon in polarisation state $\varepsilon_{j}$ due to scattering in the plane-wave background can be written (up to corrections of $\mathcal{O}(\delta n_{j}^2)$):
\be
    \delta\phi_{j}(\varphi) 
    =
    -\frac{(k^{0})^{2}}{m^{2}}\frac{1}{\eta}\int^{\vphi}  \delta n_{j}(y)~dy ;
    \quad \trm{where} \quad
    \delta n_{j}(y) 
    =
    - \frac{\varepsilon_{j}^{*} \cdot \Pi[\mathcal{A}(y)] \cdot \varepsilon_j}{2 (k_{0})^{2}}.
    \label{eqn:phaseLag2}
\ee
Now, using standard quantum field theory techniques, the polarisation operator, representing the forward scattering amplitude, can be determined to first order in perturbation theory from the Feynman diagram
\bea 
\epsilon_{j}^{\ast}\, \cdot \Pi \cdot \epsilon_{j}\, 
= \, \phantom{o}^{j}{\,\raisebox{0.4em}{\uwave{\hspace{0.5cm}}}}\TikCircle\raisebox{0.4em}{\uwave{\hspace{0.5cm}}}\!\!\phantom{o}^{j}
\label{eqn:diagPi}
\eea
where the label $j$ indicates the polarisation state of the photon. The phase lag in \eqnref{eqn:phaseLag2} has real and imaginary parts, related to the real and imaginary parts of this amplitude. The imaginary part of the amplitude for forward scattering of a photon in polarisation state $\ket{j}$ can be related to the probability of nonlinear Breit-Wheeler pair creation by the optical theorem:
\bea 
2\,\tsf{Im}\phantom{o}^{j}\raisebox{0.4em}{\uwave{\hspace{0.5cm}}}\TikCircle\raisebox{0.4em}{\uwave{\hspace{0.5cm}}}\!\!\phantom{o}^{j} = \Big|\!\!\phantom{o}^{j}\,\raisebox{0.5em}{\uwave{\hspace{0.5cm}}}\TikPair\raisebox{0.4em}\,\Big|^{2} \label{eqn:diag0}
\eea
(with the appropriate phase space factor implicitly included on the right hand side). In the exponent defining the phase lag, a positive imaginary part for $\delta n_{j}$ will manifest as a \emph{decay} of the photon state in (\ref{eqn:Adef}), corresponding to the conversion of photons into pairs.

The polarisation space of an on-shell photon propagating in the plane-wave background can be modelled as a two level system.
Without loss of generality we consider the background to have linear polarisation with electric field aligned along the $x$-direction and propagating in the $z$-direction.
The polarisation state of a counter-propagating monochromatic photon in the two-state space can be parameterised as
\begin{align}\label{eqn:PolInitial}
    \ket{\mathcal{E}_{0}}
    =
        \cos\theta_{0}
        \ket{x}
        +
        \mbox{e}^{i \psi_{0}}
        \sin\theta_{0}
        \ket{y}
    \,,
\end{align}
depending on two angles $(\theta_{0},\psi_{0})$ relative to the eigenstates of the linearly polarised background field, $\ket{x}, \ket{y}$ .
For general $\theta_{0}$ and $\psi_{0}$ the initial polarisation forms an ellipse in the two-state space with orientation defined by an angle $\tan 2\gamma_{0} = \tan 2 \theta_{0} \cos \psi_{0}$  between the ellipse's semi major-axis and $\ket{x}$\cite{Nye.rspa.1983.0109,Dennis.Opt.Comms.201.2002}.
When the relative phase angle $\psi_{0}$ is an integer multiple of $\pi$, the phase becomes real and the polarisation ellipse collapses to a line in the two-state space: the incoming photon is then linearly polarised at an angle $\theta_{0}$ to $\ket{x}$.

Now, consider the probe photon propagating through a linearly polarised plane-wave background with $e\mathcal{A}_{\mu}(\varphi) =  a_{\mu}(\varphi)$, where
\begin{align}\label{eqn:GaugePW}
    a_{\mu}(\varphi)
    =
    m\xi_{0}
    g(\varphi)
    \epsilon_{\mu}
    \,,
\end{align} 
where $\xi_{0}$ is the peak dimensionless field strength parameter, $\epsilon_{\mu}$ denotes the polarisation direction and $g(\vphi)$ is the envelope function with $|g(\varphi)|\leq 1$. The evolution of the initial polarisation of the photon through the linearly polarised plane-wave can be modelled with Jones calculus\cite{Jones.J.Opt.Soc.Am.488.1941,Hurwitz.J.Opt.Soc.Am.493.1941,Jones.J.Opt.Soc.Am.500.1941,Jones.J.Opt.Soc.Am.486.1942}.
The background field acts as a linear phase retarder $\hat{L}$ on the initial state, implying the final state
\begin{align}\label{eqn:FieldPolarisation}
    \ket{\mathcal{E}(\varphi)}
    \equiv
    \hat{L}
    \ket{\mathcal{E}_{0}}
    =
        \mbox{e}^{i \delta\phi_{x}(\varphi)}
        \cos\theta_{0}
        \ket{x}
        +
        \mbox{e}^{i \delta\phi_{y}(\varphi)}
        \mbox{e}^{i \psi_{0}}
        \sin\theta_{0}
        \ket{y}
    =
c_{x}(\varphi)
    \ket{x}
    +
    c_{y}(\varphi)
    \ket{y}
    \,.
\end{align}
For general $(\theta_{0},\psi_{0})$, the two polarisation degrees of freedom experience different vacuum refractive indices, hence different phase lags, $\delta\phi_{x} \ne \delta\phi_{y}$, and a relative phase difference develops such that $\lim\limits_{\varphi \to \infty}\ket{\mathcal{E}(\varphi)} \ne e^{i\psi}\ket{\mathcal{E}_{0}}$ for some global phase $\psi$.

The final polarisation of the photon can be characterised by mapping the complex coefficients $(c_{x},c_{y})$ to four real parameters known as the Stokes parameters, $S_{0},S_{1},S_{2},S_{3}$. 
Here, $S_{0}$ is the \emph{photon survival probability}, normalised to unity in the absence of pair creation (see below), while the remaining $S_{j}$ define the polarisation asymmetry in the linear basis $\ket{x,y}$, the superposition/rotated linear basis $\ket{1,2} = (\ket{x} \pm \ket{y})/\sqrt{2}$, and the helicity basis $\ket{+,-} = (\ket{x} \pm i \ket{y})/\sqrt{2}$, respectively. We review these Stokes parameters for a generally polarised initial photon in \cref{app:Stokes}, where we show [see \cref{S.2} and \cref{eqn:S3}] that they are are maximised when $\theta_{0} = \pm\pi/4$. 
The remaining parameter determining the ellipticity of the initial polarisation, $\psi_{0}$, is free, so can be chosen to correspond to the polarisation of photons which can be realised experimentally.
Motivated by the experimental scenario outlined in section \ref{sec:Exp}, let us focus on the case of photons produced in the helicity eigenstate $\ket{+}$, so that $\psi_{0} = \pi/2$. 
Then, normalising the Stokes parameters as $\Gamma_{j} = S_{j}/S_{0} ~~ (j \in \{1,2,3\})$, we have for $(\theta_{0}, \psi_{0}) = (\pi/4,\pi/2)$,
\begin{align}\label{eqn:AllStokes1}
    \Gamma_{1}(\varphi)
    =
    & - \tanh \left[\IM(\Delta(\varphi))\right]\,,
    \\\label{eqn:AllStokes2}
    \Gamma_{2}(\varphi)
    =
    &
    \sin\left[\RE(\Delta(\varphi))\right]
    \sech\left[\IM(\Delta(\varphi))\right]
    \,,
    \\\label{eqn:AllStokes3}
    \Gamma_{3}(\varphi)
    =
    &
    \cos\left[\RE(\Delta(\varphi))\right]
    \sech\left[\IM(\Delta(\varphi))\right]        
    \,,
\end{align}
where we have separated the phase lag into real and imaginary parts $\delta\phi_{j} = \RE \delta\phi_{j} + i \IM \delta\phi_{j}$ and employed the phase difference,
\begin{align}\label{eqn:Diff}
    \Delta(\varphi) \equiv \delta\phi_{x}(\varphi) - \delta\phi_{y}(\varphi) \,.
\end{align} 
These Stokes parameters correspond to initially $\ket{+}$ polarised photons interacting with the background field and being measured in the $\ket{x,y}$, $\ket{1,2}$, and $\ket{+,-}$ bases, respectively.
$\Gamma_{1}$ depends only on the imaginary part of the polarisation operator, and when this can be neglected is zero across the pulse.
$\Gamma_{2}$ is initially zero, since a photon with $\ket{+}$ polarisation has an equal probability of being measured in one of the $\ket{1,2}$ basis states, but then oscillates as the photon propagates through the laser.
Conversely, $\Gamma_{3}$ is initially unity since the initial polarisation is an eigenstate of the measurement basis $\ket{+,-}$. 
However, due to the plane-wave background being linearly polarised in $\ket{x,y}$, the photon picks up a relative phase difference which causes $\Gamma_{3}$ to oscillate.
In the idealised case in question, of a completely circularly-polarised, monoenergetic distribution of photons propagating through an intense plane-wave with parameters ensuring that pair creation can be neglected, the Stokes $\Gamma_{2}$ and $\Gamma_{3}$ parameters oscillate between $\pm 1$ with distance travelled through the background. 
If pairs are created, the Stokes parameters oscillate and gradually decay to zero.

To investigate the feasibility of measuring the Stokes parameters, we will use the locally constant field approximation \cite{Dinu:2013gaa} to determine the phase lags $\delta\phi_{j}$ by integrating the one-loop result for the vacuum refractive index in a constant crossed field\cite{Narozhny:1968,Baier:1975ff}:
\be
\delta n_{j} = -\frac{2\alpha}{3} \frac{m^{2}}{(k^{0})^{2}}\int_{4}^{\infty} dv \frac{1}{z}\frac{v - 4 + 3 (1 + \delta_{j,y})}{v\sqrt{v(v-4)}}
\left[\frac{i}{\pi}\int \!dt\,t\mbox{e}^{izt +it^{3}/3}\right]
, \label{eqn:Ifunc}
\ee
where $j\in\{x,y\}$, $\delta_{j,y}$ is the Kronecker delta symbol, and $z=(v/\chi)^{2/3}$ with strong-field parameter $\chi=\xi \eta$ and $\xi=|\xi_{0}g(\vphi)|$.
(Studies of the validity of the locally constant field approximation for pair creation in a plane-wave background can be found in the literature\cite{King:2019igt,Blackburn:2021cuq}.) 
The real and imaginary parts of the integral in square brackets are $\trm{Ai}'(z)$ and $\trm{Gi}'(z)$ respectively where $\trm{Gi}(z)$ is the Scorer function \cite{vallee2010airy}
\footnote{Note that in the low-$\chi$ limit, \eqnref{eqn:Ifunc} yields $\delta n_{j} \longrightarrow \alpha\chi^{2}[1+3(1+\delta_{j,y})]m^{2}/90\pi(k^{0})^2$, which is equivalent to the leading order one-loop expansion of the Heisenberg-Euler result for weak fields $E \ll m^{2}/e$ in the case of a plane-wave background, where $E$ is the plane-wave field strength.}.

\subsection{Multiple scattering in plane-wave backgrounds \label{sec:Multiple}}

The refractive index approach has been derived using the one-loop polarisation operator, to give a relative lag $\Delta \sim \mathcal{O}(\alpha)$. 
However, it is clear that an expansion of scattering amplitudes, or of the Stokes parameter, in $\Delta$ will include higher orders in $\alpha$. 
Here we consider the validity of this approach to describe multiple scattering.

An expansion in $\alpha$ of the photon-to-photon amplitude is depicted in \eqnref{eqn:diag1}, which we calculate to $\mathcal{O}(\alpha)$ (in the LCFA) through (\ref{eqn:Ifunc}). It is instructive to consider the two-loop diagrams at order $\alpha^{2}$, which include a one particle-reducible contribution with a virtual intermediate photon, as well as one particle-irreducible contributions. 
{The model used in this work exploits the fact that in the `long-pulse limit' $\alpha \xi_{0} \Phi \gg 1$, one-loop scattering of on-shell photons from the one-particle reducible term $\sim(\alpha \xi_{0} \Phi)^{2}$ dominates. Thus, when calculating the Stokes parameter, all other contributions at this order in $\alpha$ will be neglected. (See also, e.g\cite{Torgrimsson:2020gws,Fedotov:2022ely}.)
\bea 
\phantom{o}^{j}\,\raisebox{0.4em}{\uwave{\uwave{\hspace{1cm}}}}\!\phantom{o}^{j'} ~= \underbrace{\phantom{o}^{j}\,\raisebox{0.4em}{\uwave{\hspace{1cm}}}\!\phantom{o}^{j'}}_{O(\alpha^{0})} +  \underbrace{\phantom{o}^{j}\raisebox{0.4em}{\uwave{\hspace{0.5cm}}}\TikCircle\raisebox{0.4em}{\uwave{\hspace{0.5cm}}}\!\!\phantom{o}^{j'}}_{O(\alpha)}   + \underbrace{\phantom{o}^{j}\,\raisebox{0.4em}{\uwave{\hspace{0.5cm}}}\TikCircle\stackrel{\phantom{j''}}{\raisebox{0.4em}{\uwave{\hspace{0.5cm}}}}\TikCircle\raisebox{0.4em}{\uwave{\hspace{0.5cm}}}\!\phantom{o}^{j'} + \phantom{o}^{j}\raisebox{0.4em}{\uwave{\hspace{0.5cm}}}\TikCircleTwo\!\!\!\raisebox{0.4em}{\uwave{\hspace{0.5cm}}}\!\!\phantom{o}^{j'}  + \cdots}_{O(\alpha^{2})} \quad + ~~\cdots .   \label{eqn:diag1}
\eea
Then in the long-pulse limit, when there is no pair creation, the contribution to the Stokes parameter at order $\alpha^n$ involves a chain of $n$ one-loop polarisation operators where the intermediate photon is on-shell. This series is of the form of an expansion of the exponential function, with a phase dispersion altered by a vacuum refractive index differing from unity. This can be seen e.g. by solving the Schwinger-Dyson equation with a one-loop insertion into the propagator \cite{Mitter:1974yg,Meuren:2014uia}, or by solving the classical wave equation with a source term given by the one-loop diagram \cite{Bohl:2015uba}. In the Stokes parameter $\Gamma_{2}$, the $[\RE(\Delta)]^{n}$ term in the expansion of $\sin[\RE(\Delta)]$ is then taken to be, in the long-pulse limit, equivalent to the contribution to the  relative difference in phases $\delta\phi_{x}-\delta\phi_{y}$, of the $O(\alpha^n)$ chain of one-loop reducible diagrams.
Although even powers of the elastic parameter $\RE(\Delta)$ occur in $c_{\pm}$, in $\Gamma_{2}$ they turn out to cancel, and helicity flipping only occurs at odd powers of $\RE(\Delta)$ i.e. odd powers of $\alpha$. Therefore, by expanding the Stokes parameter in $\RE(\Delta)$, we can identify the contribution from single, triple, $(2n+1)$-fold photon scattering in the plane-wave background. 
By varying a suitable parameter such as the background intensity or pulse duration, one can isolate in experiment the contribution from higher-order nonlinear vacuum polarisation. 

To order the different parameter regimes, we recall the refractive index approach requires that pair creation be a small effect. 
Comparing the real and imaginary parts of the refractive index in \figrefa{fig:paramSpace}, we can see that this is satisfied for $\chi < 1$. 
Furthermore, we see that the leading-order $\chi$ expansion of the refractive index is quite accurate even up to $\chi=1$. Inserting this leading-order term in the expansion  into the definition of the relative lag parameter $\Delta$ (\eqnref{eqn:Diff}), we see that 
\begin{align}
    \lim\limits_{\chi \ll 1}
    \RE(\Delta)
    \approx
    &
    \frac{\alpha}{30\pi\eta}\int \chi^{2} d\vphi; 
    \quad
    &
    \lim\limits_{\chi \ll 1}
    \IM(\Delta) 
    \approx
    &
    \frac{\alpha}{16\eta}\sqrt{\frac{3}{2}} \int \chi \mbox{e}^{-\frac{8}{3\chi}} d\vphi 
    \,.
\end{align}
Accordingly, we use the parameters $\Delta^{\tiny\tsf{el.}} = \alpha\xi_{0}\chi_{0}\Phi/30\pi$ and $\Delta^{\tiny\tsf{inel.}} = \alpha\xi_{0}\Phi\exp(-8/3\chi_{0})$, recall (\ref{eqn:Delta.el} and (\ref{eqn:Delta.inel}), to quantify the likelihood of elastic scattering and pair-creation, respectively, where $\xi_{0}$ and $\chi_{0}$ denote the maximum values reached in a pulse.
This allows different regimes to be defined in a parameter space, illustrated in \figrefb{fig:paramSpace}.

\begin{figure}[t!!]
\centering
\includegraphics[width=0.3\linewidth]{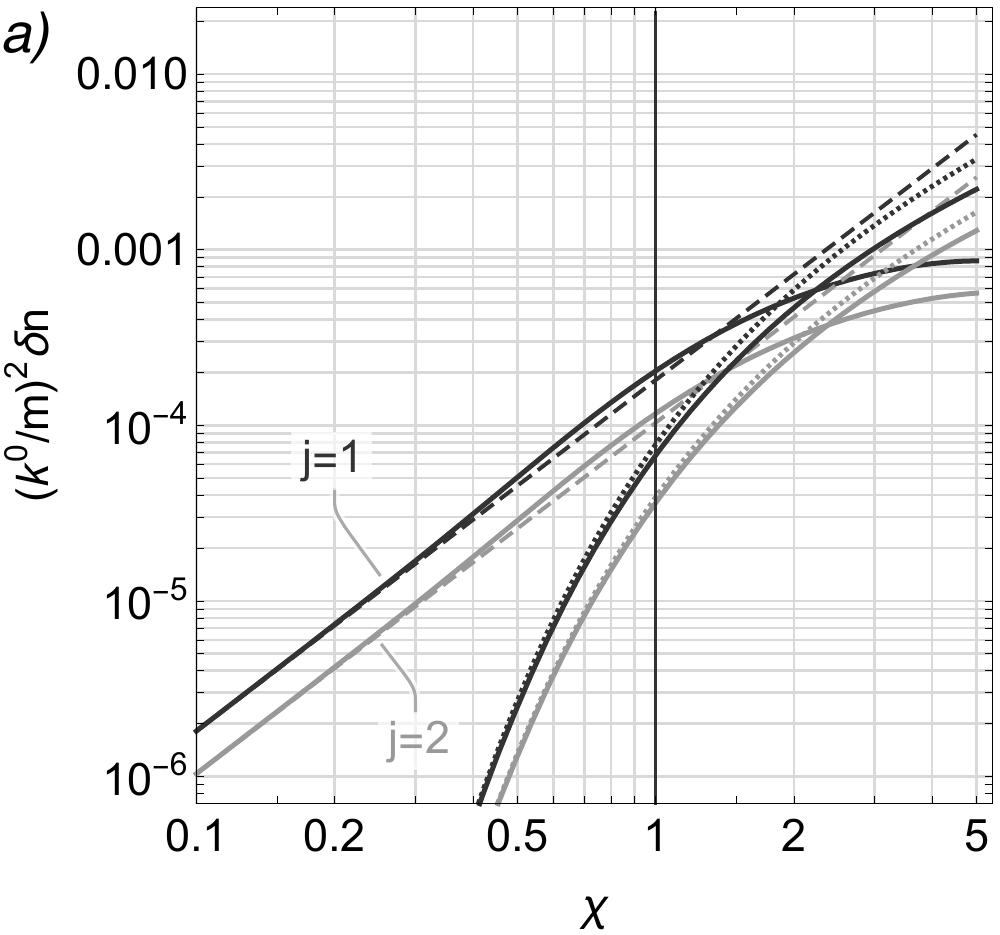}\hspace{1cm}
\includegraphics[width=0.5\linewidth]{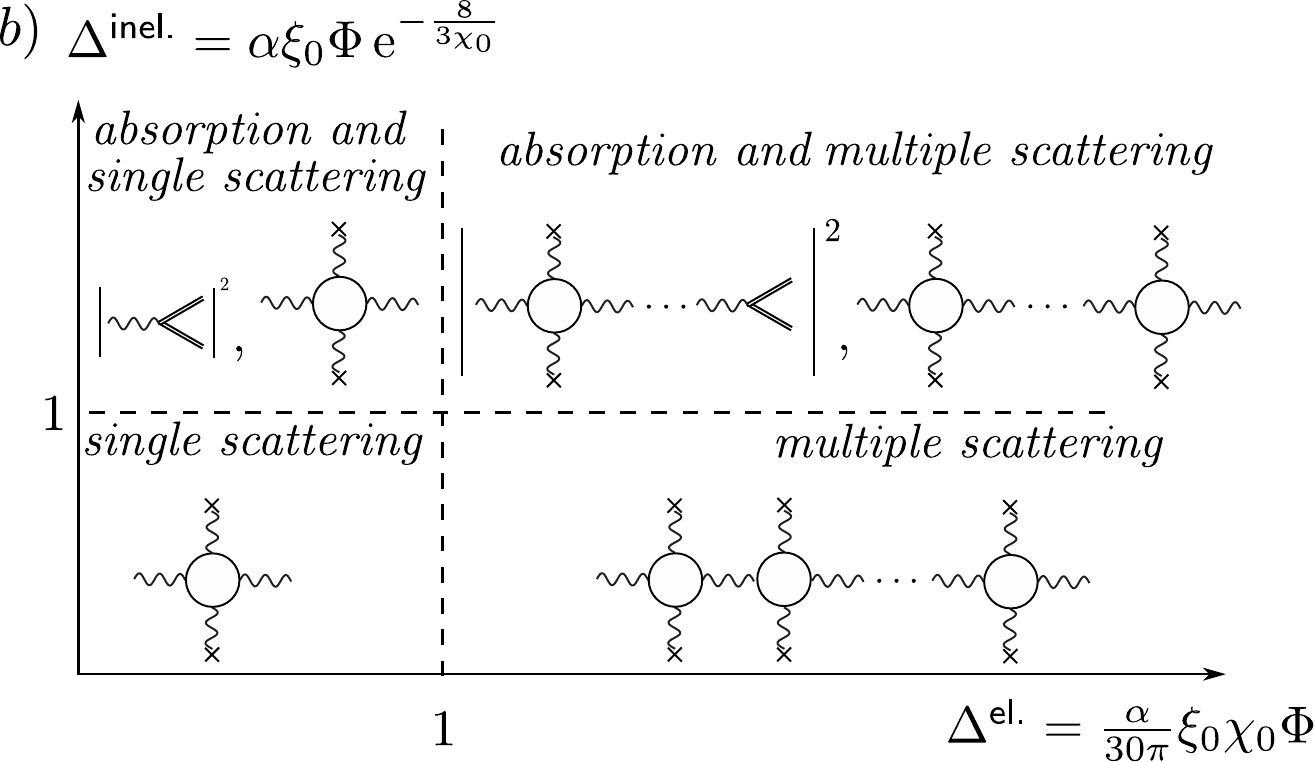}
\caption{\emph{Left}: Comparison of the vacuum refractive indices in a constant crossed field (solid lines) and the leading-order $\chi$ weak-field approximations (dashed lines). 
The real parts give a power law dependency for $\chi \ll 1$, whereas the imaginary parts are exponentially suppressed. 
\emph{Right}: Different parameter regimes of higher-order nonlinear vacuum polarisation, where $\chi_{0}= \xi_{0} \eta$.
} \label{fig:paramSpace}
\end{figure}

\section{Experimental scenario}\label{sec:Exp}

\begin{figure}
    \centering
    \includegraphics[width=\textwidth,trim={0.0cm 0.0cm 0.0cm 0.0cm},clip=true]{
        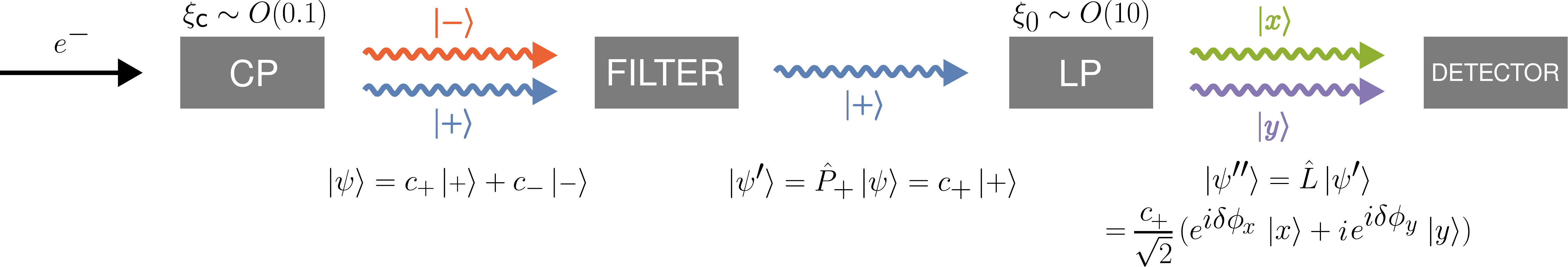
    }
    \caption{\label{fig:Schematic} 
        Schematic overview of experimental setup.
        (1) Unpolarised electrons undergo nonlinear Compton scattering off a (right-hand) circularly polarised laser pulse (CP) to produce high-energy photons in $\ket{+,-}$ states. 
        (2) Photons in the $\ket{-}$ state are filtered out by a collimator plus energy filter (operator $\hat{P}_{+}$)\cite{Tang.PhysRevA.102.022809.2020,Tang.PLB.809.135701.2020}.
        (3) Photons in the $\ket{+}$ state scatter off a long duration high-energy linearly polarised laser pulse (LP) which acts as a linear phase retarder (operator $\hat{L}$). These are then measured in a detector downstream, in the appropriate basis to reconstruct the desired Stokes parameter.
    }
\end{figure}

Measurement of vacuum birefringence with high-intensity laser systems requires production of photons with high polarisation purity. As detailed in section \ref{sec:Intro}, detecting birefringence through the phase shift of photons propagating through a secondary high-intensity laser pulse requires their initial polarisation to be complementary to the polarisation eigenstates of the background.
For example, if the secondary laser is polarised in the linear basis $\ket{x,y}$, the initial state of the photons should be either $\ket{1,2}$ or $\ket{+,-}$ polarised --- we will adopt the latter, hence helicity eigenstates.

High-energy photons of this polarisation can be produced through the scattering of high-energy electrons with lasers, often referred to as (inverse) nonlinear Compton scattering~\cite{Nikishov.JETP.1964,Brown.PR.1964,Goldman.PL.1964,Fedotov:2022ely}.
When the field strength parameter of the scattering laser, $\xi_{\tiny\tsf{c}}$, is in the weak-field regime ($\xi_{\tiny\tsf{c}} \lesssim 1$), the produced photons can be quasi-monoenergetic and highly polarised~\cite{Hartemann.PRSTAB.2005,Tang.PhysRevA.102.022809.2020,Tang.PLB.809.135701.2020}, with the polarisation state of the photons depending on the polarisation of the scattering laser.
Circular polarisation can be achieved through the collision of electrons with a circularly polarised laser pulse.
If the laser has right-hand circular polarisation,
right-hand circularly polarised photons can be produced with up to 95\% polarisation purity~\cite{Tang.PhysRevA.102.022809.2020,Tang.PLB.809.135701.2020} due to a separation of right- and left-hand polarised photons in the angular plane.

In \cref{fig:Schematic} we outline, schematically, our proposed experimental setup for measuring vacuum birefringence.
Beginning with (unpolarised) electrons\footnote{These could be sourced either from conventional accelerators or through plasma-based acceleration schemes such as laser wakefield acceleration}, we use a circular polariser consisting of a combination of a circularly polarised laser pulse, with field strength parameter $\xi_{\tiny\tsf{c}} \sim O(0.1)$, and a polarisation filter.
From the collision of the electrons with the circularly polarised laser pulse, photons are produced in both right- and left-hand polarisation modes. Complementary to section \ref{sec:Intro} we can describe their state and interaction with the lasers and optical elements using the Jones calculus. The photons produced by Compton scattering are in a superposition
 \begin{align}\label{eqn:PhotonInitialState}
     \ket{\psi}
     =
     &
     c_{+} \ket{+}
     +
     c_{-} \ket{-} \,, 
     \quad 
     &
     \ket{\pm} 
     =
     \frac{1}{\sqrt{2}} 
     \begin{pmatrix}
         1 \\ \pm i
     \end{pmatrix} 
     \,,
 \end{align}
 with $\ket{\pm}$ corresponding to right- and left-hand polarisation, respectively.

However, photons produced in the $\ket{+,-}$ states have different transverse momentum structure depending on the polarisation of the weak laser. For a right-handed circularly polarised laser pulse, high-energy photons are produced in the $\ket{+}$ state with predominantly low transverse momentum (i.e. are emitted in the electron propagation direction), while those in the $\ket{-}$ state have large transverse momentum~\cite{Tang.PhysRevA.102.022809.2020}.
Due to this angular separation, a collimator can filter out the majority of high-energy $\ket{-}$ helicity photons. (Low-energy photons produced in this state, emitted with low transverse momentum, could be filtered using an energy filter.) We thus have the action of a (lossless) circular polarisation operator, $\hat{P}_{\pm}$:
 \begin{align}\label{eqn:PhotonFiltered}
     \ket{\psi^{\prime}} 
     =
     &
     \hat{P}_{\pm} \ket{\psi} = c_{\pm} \ket{\pm} 
     \quad
     &
     \hat{P}_{\pm}
     =
     &
     \ket{\pm}\bra{\pm}
     =
     \frac{1}{2}
     \begin{pmatrix}
         1 & \mp i 
         \\
         \pm i & 1
     \end{pmatrix}
     \,,
 \end{align}
 which picks out one of the polarisation modes. Hence, $\hat{P}_{+}$ projects onto states of the form (\ref{eqn:PolInitial}) with $(\theta_{0}, \psi_{0}) = (\pi/4, \pi/2)$.

In the second stage, the polarised photons then collide with a strong laser pulse with intensity parameter $\xi_{0} \sim O(10^{2})$.
This acts as a linear phase retarder, $\hat{L}$, on the polarisation state of the photon, as the phase accumulated during propagation through the intense laser focus differs for the $\ket{x,y}$ elements of $\ket{\psi'}$:
\begin{align}\label{eqn:PohotonFinal}
     \ket{\psi^{\prime\prime}}
     =
     &
     \hat{L}
     \ket{\psi^{\prime}}
     =
     \frac{ c_{\pm}}{\sqrt{2}}
     (
     \mbox{e}^{i\delta\phi_{x}}
     \ket{x}
     \pm
     i
     \mbox{e}^{i\delta\phi_{y}}
     \ket{y}
     )
     \,,
     \quad
     &
     \hat{L}
     = \mbox{e}^{i \delta\phi_{x}} \ket{x}\bra{x} + \mbox{e}^{i \delta\phi_{y}} \ket{y}\bra{y} 
     =
     &
     \begin{pmatrix}
         \mbox{e}^{i \delta\phi_{x}} & 0 \\
         0 & \mbox{e}^{i \delta\phi_{y}}
     \end{pmatrix}
     \,.
\end{align}
This, of course, reproduces the state already given in (\ref{eqn:FieldPolarisation}). The outgoing photon is then to be measured in either the $\ket{1,2}$ or $\ket{+,-}$ polarisation bases, from which $\Gamma_{2}$ or $\Gamma_{3}$ can be determined, respectively.

The final stage thus involves the measurement of the Stokes parameter. As shown in  \cref{app:Stokes}, the Stokes parameter $\Gamma_{2}$  can be expressed as an asymmetry of probabilities, and hence an asymmetry in the number of photons $\tsf{N}_{1,2}$ detected in each polarisation state: $\Gamma_{2} = (\tsf{N}_{1} - \tsf{N}_{2}) / (\tsf{N}_{1}+\tsf{N}_{2})$. Therefore, the detector must involve some process sensitive to photon polarisation at $O(\trm{GeV})$ energies. One example is the GlueX experiment which employs a polarimeter based on the linear trident process in matter \cite{GlueX:2020idb}, and a similar scheme could be employed here.

To perform this experiment one would need access to both high-energy electrons, for the purpose of producing polarised photons, and high-intensity lasers, to induce vacuum birefringence.
Such a scheme could be possible using a combination of the L3-HAPLS and L4-ATON laser systems of the ELI Beamlines facility.
L3-HAPLS is a petawatt class Ti:Sapph ($\lambda \sim 0.815~\mu\text{m}$) laser with an output pulse energy of $> 30$~J and pulse duration of 30~fs.
This system could be utilised to produce high-energy polarised photons by splitting the beam into two pulses.
The first pulse would drive electron beam production through laser wakefield acceleration, and would consist of a significant portion of the available laser power, such that the power on target remains $P\sim 1$~PW.
This would allow the acceleration of electrons up to energies of $E_{e} \sim 10~\text{GeV}$, as has already been demonstrated experimentally with the acceleration of electrons to $E_{e} \sim 8$~GeV by a laser with power $P \approx 0.85$~PW\cite{Gonsalves:2019wnc}.
The second, lower power, pulse would then collide with the electron beam to produce polarised photons via nonlinear Compton scattering.
L4-ATON will be a multi-petawatt class Nd:glass ($\lambda \sim 1.05~\mu\text{m}$) laser system, delivering an output pulse energy of 1.5~kJ, reaching 10~PW power at a pulse duration of 150~fs.
The high output pulse energy provides a unique opportunity for the study of vacuum birefringence: We suggest to use this pulse as the linear phase retarder in \cref{fig:Schematic}, with a pulse duration stretched to the picosecond level while still maintaining $\xi_{0} \sim O(10-100)$.
This will enable not only single elastic scattering, but multiple scattering events to be observed.

\section{Monoenergetic photons on a plane-wave pulse} \label{sec:Monoenergetic}

Before considering the more realistic case of a bunch of photons with a particular energy spectrum interacting with a focussed laser pulse, we first consider the idealised case of monoenergetic photons interacting with a linearly polarised plane-wave.
This will allow us to demonstrate the key observable signatures for birefringence, in particular how multiple scattering events can be distinguished in the long pulse duration case.
We consider a linearly polarised plane-wave pulse with the gauge potential of \eqnref{eqn:GaugePW}, i.e. $e\mathcal{A}_{\mu}(\varphi) = m \xi_{0} g(\varphi) \epsilon_{\mu}$.
We model the pulse envelope as a hyperbolic secant function,
\begin{align}\label{eqn:ggaussian}
    g(\varphi)
    =
    \sin(\varphi)
    \sech\Big(
        \frac{\varphi}{\Phi}
    \Big)
    \,,
\end{align}
with phase $\varphi = \varkappa \cdot x$ and dimensionless pulse length parameter $\Phi$. 
The latter can be expressed in terms of the temporal full-width-half-maximum (fwhm) duration $\tau_{\tiny\tsf{fwhm}}$ via
\begin{align}
    \Phi = \frac{ \tau_{\tiny\tsf{fwhm}}}{\lambda \text{arcosh} 2}
    \,,
\end{align}
where $\lambda$ is the laser wavelength.
We use a hyperbolic secant function here to ensure that when we include Gaussian focussing in \cref{sec:Focus} the validity conditions of the paraxial beam approximation are satisfied\cite{mcdonald2000gaussian}.

Throughout the remainder of this paper we focus primarily on the Stokes parameter $\Gamma_{2}$, defined in \cref{eqn:AllStokes2}, and denote by $\Gamma_{2}^{(n)}$ the expansion of the Stokes parameter to $O(\Delta^{n})$,  corresponding to $(2n + 1)$-fold photon scattering.
As discussed in the introduction, multiple scattering events can be expected when the elastic scattering parameter $\Delta^{\tiny\tsf{el.}} \gg 1$, which can be achieved by increasing the field strength, $\xi_{0}$, the photon energy, $\omega$, or the pulse phase duration, $\Phi$.
However, increasing the field strength or the photon energy can quickly lead to the inelastic scattering parameter becoming large $\Delta^{\tiny\tsf{inel.}} \gg 1$ due to its nonperturbative scaling with $\chi_{0}$.
For this reason, we also introduce $\widetilde{\Gamma}_{2}$, the Stokes parameter for vanishing imaginary part, $\mathrm{Im} \, (\Delta_j) \to 0$, such that the condition $|1 - \widetilde{\Gamma}_{2}/\Gamma_{2}| \ll 1$ ensures the probability of elastic scattering is maximised without significant pair creation occurring.

\begin{figure}[t!!]
    \centering
    \includegraphics[width=0.99\textwidth,trim={0.0cm 0.0cm 0.0cm 0.0cm},clip=true]{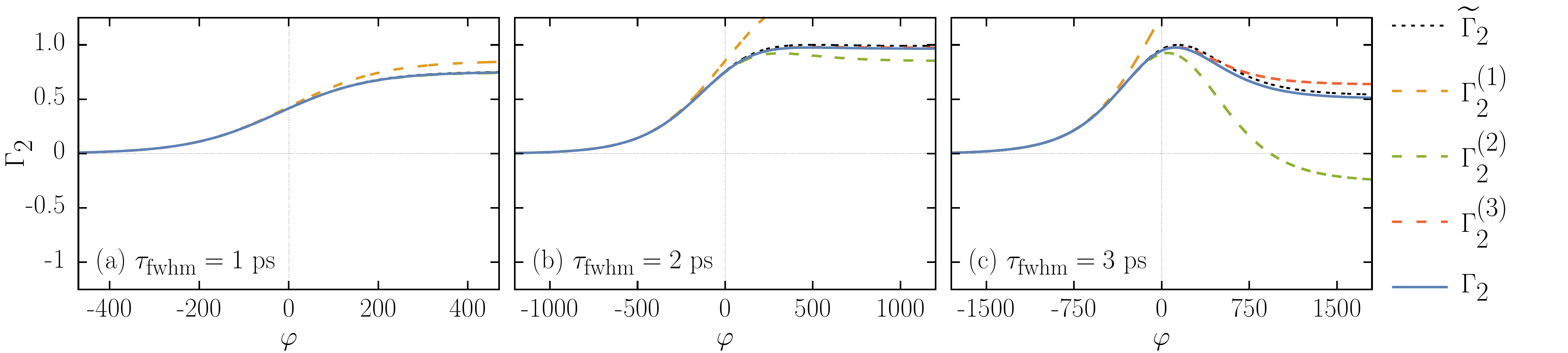}
 
    \caption{\label{fig:SingleStokes} 
        Evolution of the Stokes parameter, $\Gamma_{2}$, with the laser phase, $\varphi$, for a single photon of energy $\omega = 2$~GeV in a linearly polarised plane-wave field with $\xi_{0} = 50$, wavelength $\lambda = 1.05$~$\mu\text{m}$, and full-width-half-maximum pulse durations, $\tau_{\tiny\tsf{fwhm}}$: (a) 1~ps, (b) 2~ps, (c) 3~ps.
        Solid: all-orders Stokes parameter $\Gamma_{2}$ with pair creation.
        Dashed: $(2n + 1)$-fold photon scattering with $n = 1$ (orange), $n = 2$ (green), and $n = 3$ (red). 
        Black dot: all-orders Stokes parameter $\widetilde{\Gamma}_{2}$ without pair creation.
    }
\end{figure}

In \cref{fig:SingleStokes} we consider the scattering of monoenergetic photons of energy $\omega = 2$~GeV with a linearly polarised plane-wave field with dimensionless intensity parameter $\xi_{0} = 50$ and wavelength $\lambda = 1.05$~$\mu\text{m}$ for three different durations of the laser pulse: $\tau_{\tiny\tsf{fwhm}} = (1,2,3)$~ps.
In each case we show the evolution of: the all-orders Stokes parameter with pair creation included, $\Gamma_{2}$ (blue, solid); the all-orders Stokes parameter without pair creation (black, dot-dashed), $\widetilde{\Gamma}_{2}$; and the $O(\Delta^{n})$ contributions to the Stokes parameter for $n=1$ (orange, dashed), $n=2$ (green, dashed), and $n=3$ (red, dashed).

The number of scattering events which occur as the photon interacts with the laser pulse is determined by comparing the all-orders Stokes parameter to the $O(\Delta^{n})$ contributions.
When the all-orders Stokes parameter is well approximated by the $n^{\tiny\tsf{th}}$-order perturbative expansion, we can reasonably deduce that the photon has undergone at least $(2n + 1)$-fold photon scattering.
By considering the elastic scattering parameter, $\Delta^{\tiny\tsf{el.}}$, which takes values between $2 < \Delta^{\tiny\tsf{el.}} < 6$ for the parameters in \cref{fig:SingleStokes}, one expects that multiple scattering of the photon should be observed.
For $\tau_{\tiny\tsf{fwhm}} = 1$~ps, one can distinguish the $O(\Delta^{1})$ term $\Gamma_{2}^{(1)}$ from the all-orders Stokes parameter with a relative difference in the final values of order $|1 - \Gamma_{2}^{(1)}/\Gamma_{2}| \approx 0.13$, and the evolution of $\Gamma_{2}$ is well approximated by including terms up to $O(\Delta^{2})$ in the perturbative expansion, with $|1 - \Gamma_{2}^{(2)}/\Gamma_{2}| \approx 0.005$.
This suggests that the photon has undergone at least $5$-fold photon scattering in the laser pulse.
Increasing the duration up to $\tau_{\tiny\tsf{fwhm}} = 2$~ps, the difference to the $O(\Delta^{2})$ perturbative expansion inreases, with $|1 - \Gamma_{2}^{(2)}/\Gamma_{2}| \approx 0.11$, and we instead find that the all-orders result is well approximated by terms up to $O(\Delta^{3})$ with $|1 - \Gamma_{2}^{(3)}/\Gamma_{2}| \approx 0.008$.
The longest duration of $\tau_{\tiny\tsf{fwhm}} = 3$~ps requires terms up to order $O(\Delta^{4})$ to be included (not shown in figure) to accurately model the Stokes parameter evolution, with $|1 - \Gamma_{2}^{(4)}/\Gamma_{2}| \approx 0.02$, such that the photon has undergone at least $9$-fold scattering.
We thus see that the parameter $\Delta^{\tiny\tsf{el.}}$ provides a useful qualitative estimate for the onset of multiple scattering. 

Across the range of pulse durations considered in \cref{fig:SingleStokes}, the evolution of the Stokes parameter is not affected by the inclusion of pair creation apart from a small effect in the longest duration case.
This is not unexpected considering the magnitude of $\Delta^{\tiny\tsf{inel.}}$, which for $\tau_{\tiny\tsf{fwhm}} = 1$~ps takes the value $\Delta^{\tiny\tsf{inel.}} \approx 0.2$, increasing to $\Delta^{\tiny\tsf{inel.}} \approx 0.6$ for $\tau_{\tiny\tsf{fwhm}} = 3.0$~ps. Nevertheless, the discrepancy between the final values of the Stokes parameters with and without pair creation remains small even at $\tau_{\tiny\tsf{fwhm}} = 3$~ps, when $|1 - \widetilde{\Gamma}_{2}/\Gamma_{2}| \approx 0.06$.
However, to properly assess the importance of pair creation for different choices of parameters, we should compare cases in which the elastic scattering parameter, $\Delta^{\tiny\tsf{el.}}$, is kept constant while the inelastic parameter, $\Delta^{\tiny\tsf{inel.}}$, is allowed to vary.
As $\Delta^{\tiny\tsf{el.}} \propto \xi_{0}^{2} \Phi$, doubling the field strength while quartering the duration will keep $\Delta^{\tiny\tsf{el.}}$  constant. $\Delta^{\tiny\tsf{inel.}}$, however, depends nonperturbatively on $\xi_{0}$, so doubling the field strength can drastically increase $\Delta^{\tiny\tsf{inel.}}$ and hence the probability of pair creation.
This is demonstrated in \cref{fig:PairsComparison}.
In each sub-figure we use a monoenergetic photon of energy $\omega = 2$~GeV, and the black dot-dashed lines correspond to neglecting pair creation.
The blue solid lines use a field strength parameter of $\xi_{0} = 50$, and pulse durations: (a) $\tau_{\tiny\tsf{fwhm}} = 1$~ps, (b) $\tau_{\tiny\tsf{fwhm}} = 2$~ps, (c) $\tau_{\tiny\tsf{fwhm}} = 3$~ps.
We see that as we increase the pulse duration, the relative difference between including and neglecting pair creation remains small, due primarily to the inelastic scattering parameter remaining $\Delta^{\tiny\tsf{inel.}} \ll 1$, with only a small discrepancy in the longest pulse duration case as then $\Delta^{\tiny\tsf{inel.}} \approx 1$.
This is contrasted with the red dashed lines which use a field strength parameter of $\xi_{0} = 100$, and pulse durations: (a) $\tau_{\tiny\tsf{fwhm}} = 250$~fs, (b) $\tau_{\tiny\tsf{fwhm}} = 500$~fs, (c) $\tau_{\tiny\tsf{fwhm}} = 750$~fs.
Doubling the field strength leads to substantially greater values of the inelastic scatering parameter, which becomes $\Delta^{\tiny\tsf{inel.}} \approx 0.86$ in (b), and as a result there is a clear decay of the Stokes parameter.
This could, perhaps, be used to optimise an experiment to minimise the probability of pair creation occurring, or to tailor the number of scattering events which one wishes to observe to a particular number.
In the remainder of this paper we choose to optimise parameters to minimise pair creation.

\begin{figure}[t!!]
    \centering
    \includegraphics[width=0.99\textwidth,trim={0.0cm 0.0cm 0.0cm 0.0cm},clip=true]{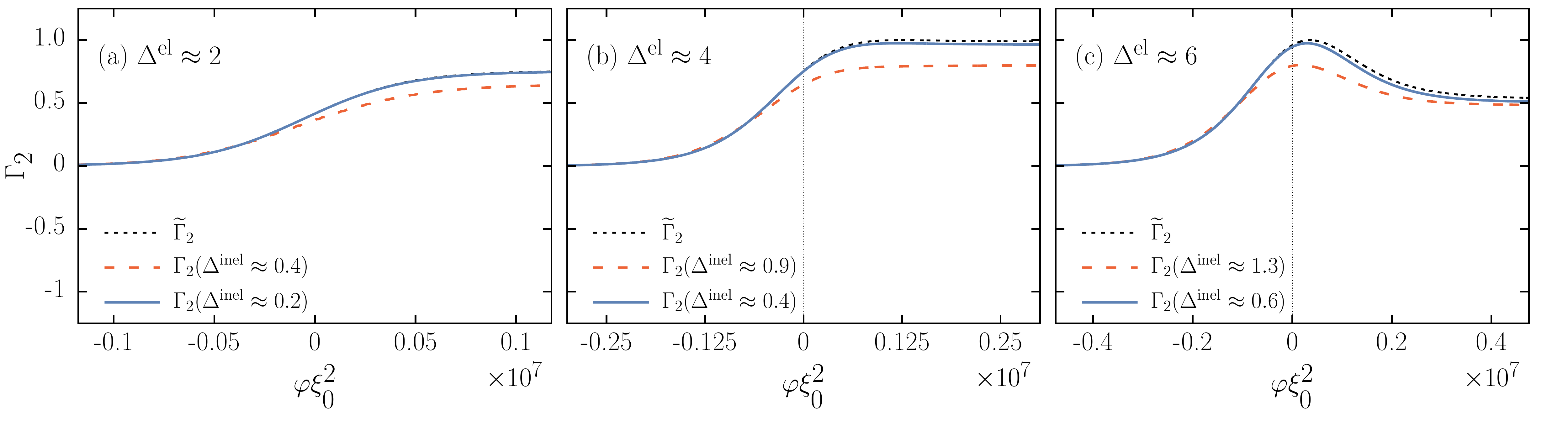}
 
    \caption{\label{fig:PairsComparison} 
      Evolution of the Stokes parameter, $\Gamma_2$, with the laser phase, $\varphi$, for different $\Delta^{\mathrm{inel}}$ at constant $\Delta^{\mathrm{el}}$.
      Probe photons have energy $\omega = 2$~GeV.
      \emph{Dotted lines (black)}: All-orders Stokes parameter neglecting pair creation.
      \emph{Solid lines (blue)}: Ditto with pair creation included, $\xi_{0} = 50$ and pulse durations: (a) $\tau_{\tiny\tsf{fwhm}} = 1$~ps, (b) $\tau_{\tiny\tsf{fwhm}} = 2$~ps, (c) $\tau_{\tiny\tsf{fwhm}} = 3$~ps.
      \emph{Dashed lines (red)}: Ditto with pair creation included, $\xi_{0} = 100$ and pulse durations: (a) $\tau_{\tiny\tsf{fwhm}} = 250$~fs, (b) $\tau_{\tiny\tsf{fwhm}} = 500$~fs, (c) $\tau_{\tiny\tsf{fwhm}} = 750$~ps.
    }
\end{figure}

\section{Photon distributions on plane-wave backgrounds} \label{sec:EnergySpectra}

Having demonstrated the key experimental signature for birefringence and multiple photon scattering in the idealised case, it is important to understand how the signal will be affected by switching to more realistic experimental parameters. In a first step, we generalise the analysis to non-monoenergetic photon sources, thus allowing for a frequency distribution. Leaving the phase dependence implicit, the Stokes parameter, $\Gamma_{2}(\omega)$, for a single photon with energy $\omega$ is given by \cref{eqn:AllStokes2}. As highlighted in \cref{app:Stokes}, this can be expressed as an asymmetry of probabilities, $\Gamma_{2}(\omega) = [\prob_{1}(\omega) - \prob_{2}(\omega)] / [\prob_{1}(\omega) + \prob_{2}(\omega)]$. 
For a photon bunch with energy distribution $\mathcal{N}(\omega)$, the number of photons measured in the $\ket{i = 1,2}$ basis will be
\begin{align}\label{eqn:Number}
    \mathsf{N}_{i}
    =
    \int_{0}^{\infty} \ud \omega
    \,
    \mathcal{N}(\omega) 
    \prob_{i}(\omega)
    \, , \quad i = 1,2 \; .
\end{align}
The final Stokes parameter will thus be an \emph{average} over a distribution, and can be determined through the asymmetry,
\begin{align}\label{eqn:StokesN}
    \braket{\Gamma_{2}}
    =
    \frac{\mathsf{N}_{1} - \mathsf{N}_{2}}{\mathsf{N}_{1} + \mathsf{N}_{2}}
    \,,
\end{align}
where the angle brackets denote the distributional average. More explicitly, the averaged Stokes parameter is
\begin{equation}\label{eqn:StokesDistribution}
    \braket{\Gamma_{2}}
    =
    \frac{
        \int_{0}^{\infty} \ud \omega
        \,
        \mathcal{N}(\omega) 
        \sin\left[\RE(\Delta(\omega))\right]
    }{
        \int_{0}^{\infty} \ud \omega
        \,
        \mathcal{N}(\omega) 
    \cosh\left[\IM(\Delta(\omega))\right]}
    \,.
\end{equation}
Since the photon distribution $\mathcal{N}(\omega)$ enters in both the numerator and denominator of \cref{eqn:StokesDistribution}, the final value of $\braket{\Gamma_{2}}$ is insensitive to the overall normalisation of the spectrum, and hence to the total number of photons which interact with the laser\footnote{We note, however, that a larger number of photons will improve the overall statistics in an experiment, and so photon sources which produce large numbers of photons are beneficial.}.
In the next subsections, the following averaged Stokes parameters will be employed: $\braket{\Gamma_{2}}$ and  $\braket{\widetilde{\Gamma}_{2}}$ denote the all-order results with and without pairs, respectively; $\braket{\Gamma_{2}^{(n)}}$ corresponds to the perturbative expansion to $O(\Delta^n)$ and  $\Gamma_{2}(\omega)$ is the the all-order value for a single photon with energy $\omega$.

\subsection{Gaussian distribution} \label{sec:EnergySpectraGaussian}

\begin{figure}[t!!]
    \centering
    \includegraphics[width=0.99\textwidth,trim={0.0cm 0.0cm 0.0cm 0.0cm},clip=true]{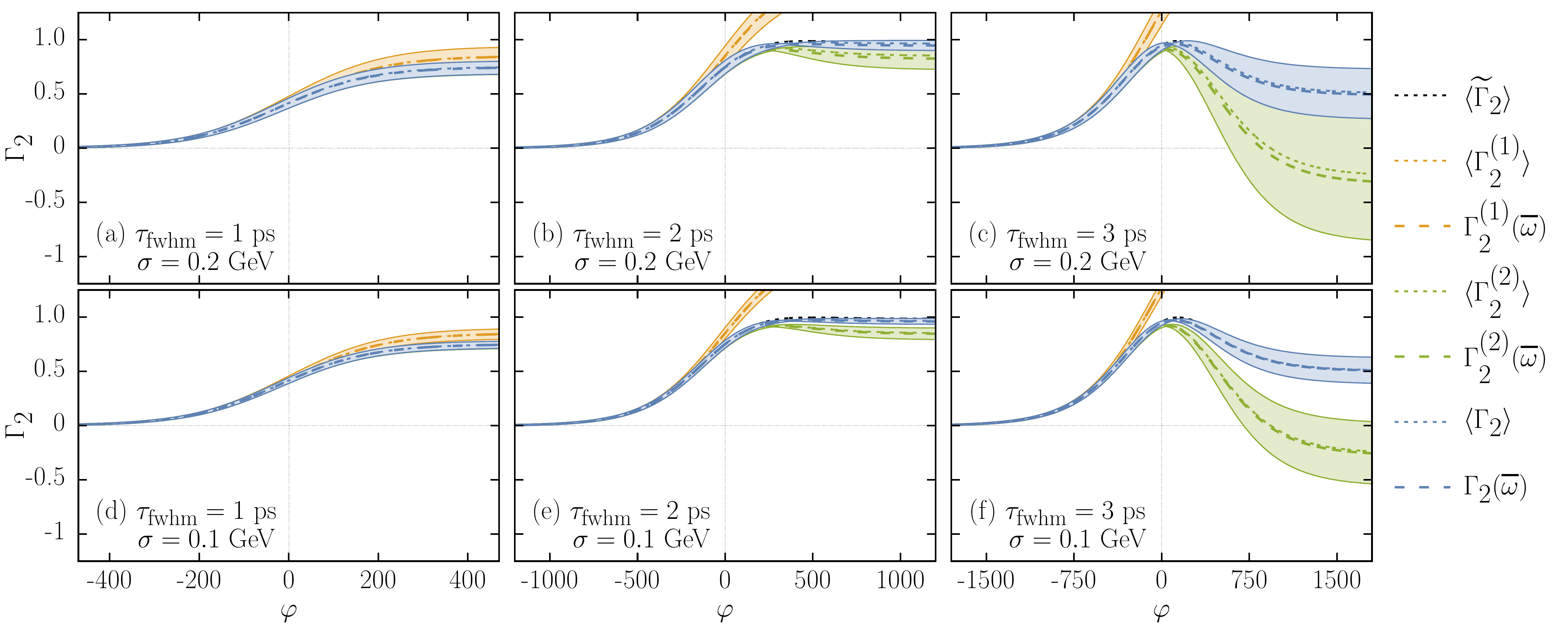}
    
    \caption{\label{fig:GaussianStokes} 
        Evolution of the Stokes parameter, $\Gamma_2$, for photons with Gaussian energy spectrum (mean energy $\overline{\omega} = 2$~GeV, standard deviation $\sigma = 0.2$~GeV (top row) or $\sigma = 0.1$~GeV (bottom row)) scattering off a linearly polarised laser with parameters as in \cref{fig:SingleStokes}.
       \emph{Black dot lines}: all-orders averaged Stokes parameter neglecting pair creation, 
       \emph{Dashed lines}: ditto with pair creation included to all orders (blue), $O(\Delta^{1})$ (orange), $O(\Delta^{2})$ (green), $O(\Delta^{3})$ (red).
       \emph{Dotted lines}: Stokes parameter for a single monoenergetic photon with $\omega = \overline{\omega}$.
       \emph{Solid lines}: Stokes parameter for single photons with $\omega = \overline{\omega} \pm \sigma$, bounding shaded 1~$\sigma$ bands.
    }
\end{figure}

A non-monoenergetic source, by definition, contains photons with different energies.
This means that each photon in the distribution will have a different elastic scattering parameter, $\Delta^{\tiny\textsf{el.}}$, and as a result will undergo different numbers of scattering events. (For the regime we are mostly considering, i.e.\ $\chi\lesssim1$, low-energy photons scatter less frequently than high-energy photons, but for higher values of $\chi$, there is a regime of anomalous dispersion where this trend can be reversed.) To highlight the spread of scattering parameters, we first consider an idealised symmetric energy distribution such as a Gaussian, 
\begin{align}\label{eqn:Gaussian}
    \mathcal{N}(\omega)
    =
    \exp\Big[- \frac{(\omega - \overline{\omega})^{2}}{2 \sigma^{2}}\Big] 
    \,,
\end{align}
where $\sigma^{2}$ and $\overline{\omega}$ denote variance and mean photon energy, respectively. We have omitted the overall normalisation as it cancels in the distribution-averaged Stokes parameter, cf.~\cref{eqn:StokesN,eqn:StokesDistribution}.

In \cref{fig:GaussianStokes} we show the evolution of the averaged Stokes parameter for a Gaussian frequency distribution with mean photon energy $\overline{\omega} = 2$~GeV, and two different values of the standard deviation, $\sigma = (0.1,0.2)$~GeV, corresponding to 5\% and 10\% of the mean photon energy, respectively.
In each plot the dashed (dot-dashed) line represents the all-orders Stokes parameter with (without) pair creation included to all orders (blue), $O(\Delta^{1})$ (orange), $O(\Delta^{2})$ (green) and $O(\Delta^{3})$ (red).
We will often refer to the perturbative contributions, $O(\Delta^n)$, as leading order (LO, $n=1$), next-to-leading order (NLO, $n=2$) and so forth. The dotted lines (same colour designation) correspond to the Stokes parameter for a single monoenergetic photon with energy $\omega = \overline{\omega}$, while the shaded band is bounded by the Stokes parameter for single photons with energies one standard deviation from the mean, $\omega = \overline{\omega} \pm \sigma$.

As before, we vary the pulse duration and observe the effect this has on the evolution of the Stokes parameter.
For the lower duration of $\tau_{\tiny\textsf{fwhm}} = 1$~ps, we see a separation of the (averaged) all-orders Stokes parameter and the first-order perturbative result. For both values of $\sigma$, the final value of the averaged Stokes parameter is well approximated by that of a single monoenergetic photon with energy at the spectral mean.
The bands corresponding to $\omega = \overline{\omega} \pm \sigma$ (in both the perturbative and all-orders case) are initially very narrow, but begin to diverge from the mean as the photons evolve in phase, with the higher variance case leading to a larger spread in the final Stokes parameter.
Increasing the duration to $\tau_{\tiny\textsf{fwhm}} = 2$~ps we see a further separation between the all-orders result and the LO and NLO perturbative approximants. One has to include the third order in perturbation theory (NNLO), or $O(\Delta^{3})$, to achieve a good agreement with the all-orders result.
At larger pulse duration, the difference in width of the distributions begins to become more apparent: for the narrower distribution ($\sigma = 0.1$~GeV), the averaged all-orders and perturbative results are well approximated by the single monoenergetic photon scenario at mean energy, and the relative spread of the $\pm \sigma$ bands remains quite narrow.
However, the crossing of the $\pm \sigma$ bands just after the peak of the laser pulse (in the all-orders and $O(\Delta^{2})$ case) clearly indicates the energy dependence in the evolution of the Stokes parameter.
Increasing the standard deviation of the distribution to $\sigma = 0.2$~GeV, the relative spread of the Stokes parameters  grows significantly. The averaged Stokes parameters, both at low and all orders, begin to deviate from the monoenergetic result.
This becomes more significant when the pulse duration is increased to $\tau_{\tiny\textsf{fwhm}} = 3$~ps.

\subsection{Nonlinear Compton photons} \label{sec:EnergySpectraNLC}

\begin{figure}[t!!]
    \centering
    \includegraphics[width=0.80\textwidth,trim={0.0cm 0.0cm 0.0cm 0.0cm},clip=true]{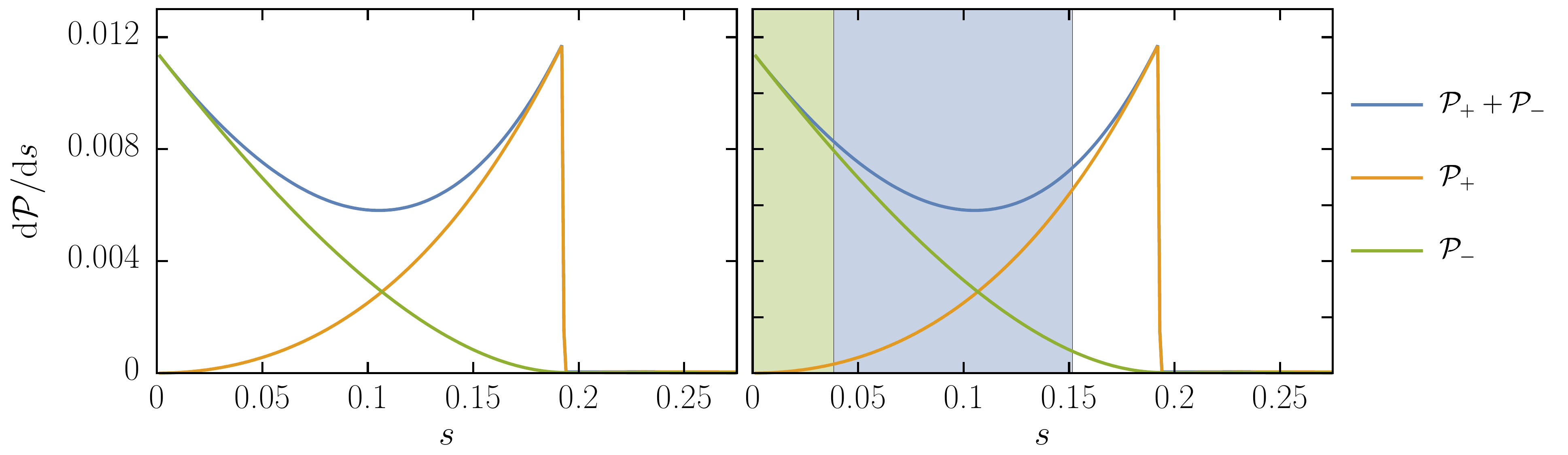}
    
    \caption{\label{fig:NLCSpectra} 
        \emph{Left}: Spectra for polarised photons produced by nonlinear Compton scattering between 10~GeV electrons and a circularly polarised laser background (strength $\xi_{\tiny\tsf{c}} = 0.1$).
        The blue line corresponds to the total differential probability, $\ud \mathcal{P}/\ud s = \ud \mathcal{P}_{+}/\ud s + \ud \mathcal{P}_{-}/\ud s$, which is the sum of individual modes polarised in directions $\varepsilon_{+}$ and $\varepsilon_{-}$ (orange and green lines, respectively). 
        \emph{Right}: The shaded blue region represents photons which would be removed by a transverse momentum cut of $r_{\tiny\tsf{cut}}^{2} = (8/25)\eta_{e} s_{\star,1}$ (see text). The shaded green region depicts the remaining low-energy, forward-scattered photons which would be removed by an energy filter.
    }
\end{figure}

One of the key issues for measuring polarisation/helicity flips is the ability to first produce polarised high-energy photons in the lab. A standard method, originally suggested in the 1960s\cite{Milburn:1962jv,Arutyunyan1964}, is to employ (inverse) Compton scattering of laser photons and high-energy electrons. (For an early experimental review, see e.g. \cite{Babusci:1990} and references therein.)
Following the recent works\cite{Tang.PhysRevA.102.022809.2020,Tang.PLB.809.135701.2020,Tang.arXiv.2022}, we consider polarised photons produced through the interaction of high-energy electrons with a circularly polarised plane-wave field, with gauge potential,
\begin{align}\label{eqn:GaugeC}
    a_{\tiny\tsf{c}}^{\mu}(\varphi)
    =
    &
    \frac{1}{\sqrt{2}}
    m
    \xi_{\tiny\tsf{c}}
    f(\varphi)
    \big(
        \varepsilon_{+}^{\mu}
        e^{-i\varphi}
        +
        \varepsilon_{-}^{\mu}
        e^{+i\varphi}
    \big)
    \,,
    \quad&
    f(\varphi)
    =
    &
    \,
    \bigg\{
    \,
    \begin{matrix}
        \sin^{2}\big(\frac{\varphi}{2N}\big) & 0 < \varphi < 2 N \pi \\
        0 & \text{elsewhere} \\
    \end{matrix}
    \;.
\end{align}
This has right-handed polarisation, $\varepsilon_{+} = (\varepsilon_{x} + i \varepsilon_{y})/\sqrt{2}$, and for completeness left-handed polarisation is defined by $\varepsilon_{-} = (\varepsilon_{x} - i \varepsilon_{y})/\sqrt{2}$.
To model the spectra of the produced photons, we calculate nonlinear Compton scattering following~\cite{Tang.arXiv.2022} in the locally monochromatic approximation~\cite{Heinzl.PRA.2020}.
We denote by $\mathcal{P}_{+}$ ($\mathcal{P}_{-}$) the probability for photons with right-handed (left-handed) polarisation, corresponding to polarisation vectors $\varepsilon_{+}$ ($\varepsilon_{-}$). As shown in~\cite{Tang.PhysRevA.102.022809.2020}, low-energy photons are predominantly produced with left-handed polarisation $\varepsilon_{-}$, while high-energy photons are predominantly produced with right-handed polarisation, $\varepsilon_{+}$.
The sharpness of the peak in the differential probability $\ud\mathcal{P}_{+}/\ud s$ depends on the field strength, $\xi_{\tiny\tsf{c}}$, since higher harmonics become important when $\xi_{\tiny\tsf{c}} \gtrsim 1$.
Regarding the choice of parameters, we recall that accelerating electrons up to 8~GeV via laser wakefield acceleration has already been demonstrated using a 0.85~PW laser system~\cite{Gonsalves:2019wnc,Kim:2021jjy}, and could in principle be achieved using the ELI Beamlines L3-HAPLS laser system.
In \cref{fig:NLCSpectra}, we thus consider the photon spectra produced by monoenergetic electrons of energy $E_{e} = 10$~GeV interacting with a circularly polarised laser with field strength parameter $\xi_{\tiny\tsf{c}} = 0.1$ and $N = 16$ cycles.
We focus on the weak-field case, $\xi_{\tiny\tsf{c}} = 0.1$, as an increase in field strength leads to spectral broadening due to the field induced shifting of the harmonic range, and this leads to a subsequent decrease in the polarisation purity in the parallel polarised mode (alternative schemes employing a linear chirp to counteract spectral broadening at high intensities have been suggested \cite{Ghbregziabher:2012xy,Terzic:2013ysa, Seipt:2014yga}).

To achieve the highest purity of polarised photons, we need to have some method for removing the photons produced in the unwanted polarisation mode. With this in mind, we recall that the locally monochromatic approximation of nonlinear Compton scattering for the emission of a photon with momentum $l_{\mu}$ from an electron with momentum $p_{\mu}$ can be written as a differential rate\cite{Heinzl.PRA.2020} per unit photon perpendicular momentum, $l_{\perp}$. 
This leads to a kinematic relationship with the lightfront momentum fraction $s = \varkappa \cdot l/\varkappa \cdot p$, which at local field strength $\xi=\xi(\vphi)$ takes the form:
\begin{align}\label{eqn:MomentumCondition}
    |l_{\lcperp} - s p_{\lcperp}|^{2}
    =
    &
    m^{2}
    \left[
        2 \eta_{e} (1 - s) s n
        -
        s^{2}
        (1 + \xi^{2})
    \right] 
    \,,
\end{align}
where $p_{\lcperp}$ is the perpendicular momentum of the incoming electron, $\eta_{e} = \varkappa \cdot p/m^{2}$, with $m$ the electron mass, and $n$ is the harmonic number.
Choosing $p_{\lcperp} = 0$, the perpendicular momentum of the emitted photon has magnitude $|l_\perp| \equiv mr$, where
\begin{align}\label{eqn:MomentumConditionR}
    r^2
    =   2 \eta_{e} n
        \frac{s(s_{\star,n} - s)}{s_{\star,n}}
    \quad \mbox{with}
    \quad
    s_{\star,n}
    =
    \frac{2 \eta_{e} n}{1 + 2 \eta_{e} n + \xi^{2}
    }
    \,.
\end{align}
We note that $r$ defines the photon emission angle, $\vartheta$, counted from the initial electron propagation direction, $r \approx (\omega/m)\sin\vartheta$, where $\omega$ is the photon energy. From (\ref{eqn:MomentumConditionR}), the value of $s$ for a given harmonic is bounded by $0 < s < s_{\star,n}$. The boundary values are the roots of $r$ which reaches a maximum at the midpoint, $s = s_{\star,n}/2$.  Thus, detecting only photons with momenta $0 < r < r_{\tiny\tsf{cut}}$ for some cut-off $r_{\tiny\tsf{cut}}$, yields two conditions on $s$, one for each half of the total interval, corresponding to the green and non-shaded regions in the right-hand plot of \figref{fig:NLCSpectra} respectively:
\begin{align}\label{eqn:Bound1}
    0
    <
    s
    <
    \frac{1}{2}
    s_{\star,n}
    -
    \frac{1}{2}
    \sqrt{
        s_{\star,n}
        \bigg(
            s_{\star,n}
            -
            \frac{2 r_{\tiny\tsf{cut}}^{2}}{n \eta}
        \bigg) 
    }
    \quad
    \text{and}
    \quad
    \frac{1}{2}
    s_{\star,n}
    +
    \frac{1}{2}
    \sqrt{
        s_{\star,n}
        \bigg(
            s_{\star,n}
            -
            \frac{2 r_{\tiny\tsf{cut}}^{2}}{n \eta}
        \bigg) 
    }
    <
    &
    s
    <
    s_{\star,n}
    \,.
\end{align}
A transverse-momentum cut can be realised with a simple collimator which will select photons from the two bins  (\ref{eqn:Bound1}) with small and large $s$. As $s \approx (\omega/2E_{e})(1+\cos\vartheta)$, the limit $s \to 0$ corresponds to either low-energy photons (which include photons emitted in the propagation direction of the electrons)
 or higher energy photons emitted in the laser propagation direction. Hence, the low-$s$ photons passing through the collimator will have low energy compared to the photons from the high-$s$ bin. Combining the collimator with an energy filter absorbing the low-energy photons will thus lead to a very high polarisation purity of the remaining high-energy photons, as demonstrated in~\cite{Tang.PLB.809.135701.2020}.
At low intensity, $\xi_{\tiny\tsf{c}} = 0.1$, the leading $n = 1$ harmonic is the dominant contribution to the photon spectrum, and the polarisation purity in the parallel mode ($\varepsilon_{+}$) for photons with $s > 0.8 s_{\star,1}$ is about 96\%.
Choosing $r_{\tiny\tsf{cut}}^{2} = (8/25) \eta_{e} s_{\star,1}$ in \cref{eqn:Bound1}, where the local quantity $\xi(\vphi)$ is replaced with its maximum value, would sufficiently filter out all Compton photons with $s_{\star,1}/5 < s < 4 s_{\star,1}/5$, see the blue shaded region in the right hand plot of \cref{fig:NLCSpectra}.
According to the arguments presented above, the remaining low-$s$ photons could then be removed by a simple energy filter represented by the green shaded region of \cref{fig:NLCSpectra}.

Having gained confidence that (mildly) nonlinear Compton photons can be produced with high polarisation purity, we move on to using them as polarised probes interacting with a linearly polarised plane-wave as given in \cref{eqn:GaugePW}.
We focus on photons polarised parallel to the background field, i.e., right-hand circularly polarised photons with polarisation vector $\varepsilon_{+}$.
Following the previous discussion we compare the evolution of the Stokes parameter employing Compton photons both with and without spectral filtering (in energy and transverse momentum). Hence, we compare the Stokes parameter averaged over the full $\mathcal{P}_{+}$ spectrum (\cref{fig:NLCSpectra}, left plot, orange line) with the truncated spectrum after collimation and energy filtering (\cref{fig:NLCSpectra}, right plot, orange line restricted to white region). In reality, the latter case would involve a small number of photons with polarisation $\varepsilon_{-}$ from the right-hand tail under the green line in \cref{fig:NLCSpectra}, but we neglect these to highlight how cutting out the low-$s$ tail in the $\mathcal{P}_{+}$ spectrum affects the evolution of the Stokes parameter.

\begin{figure}[t!!]
    \centering
    \includegraphics[width=0.99\textwidth,trim={0.0cm 0.0cm 0.0cm 0.0cm},clip=true]{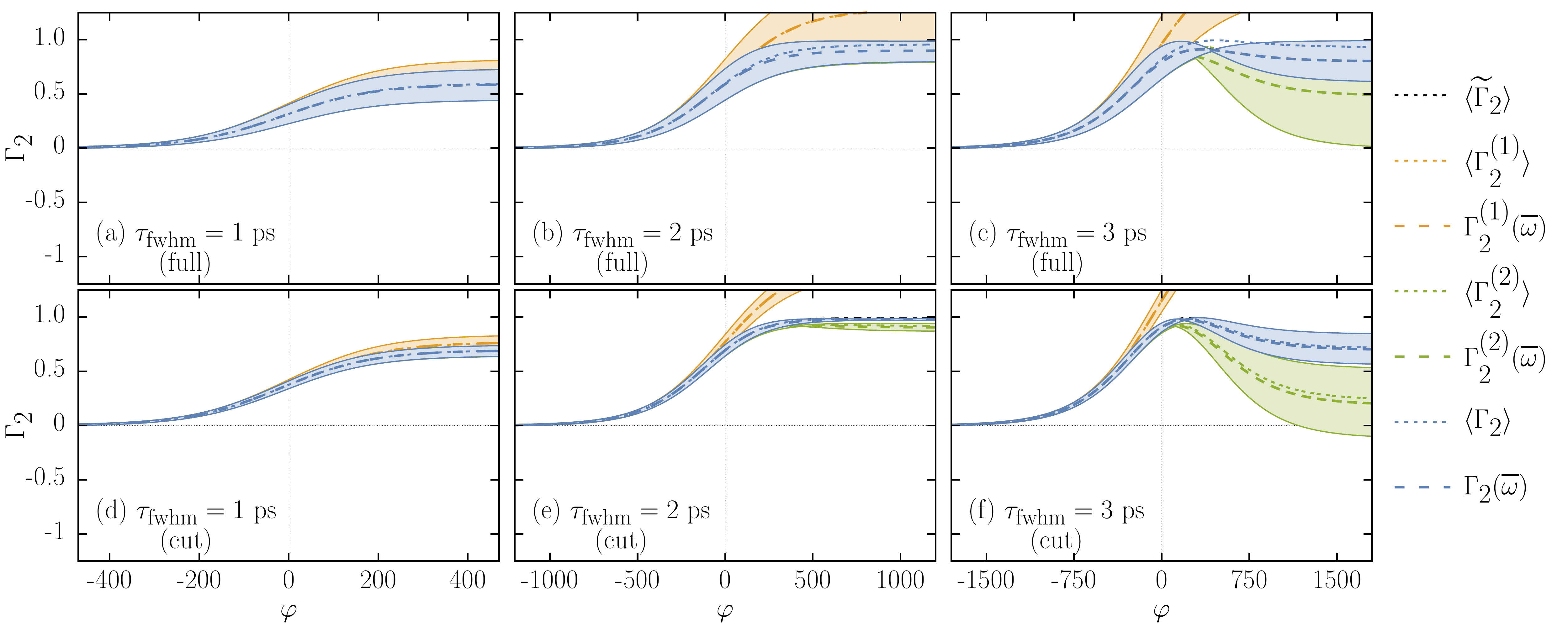}
    
    \caption{\label{fig:NLCStokes} 
        Evolution of the Stokes parameter, $\Gamma_2$, with the laser phase, $\varphi$, for probe photons produced by Compton scattering (between 10~GeV electrons and a weak circularly polarised laser pulse with $\xi_{\tiny\tsf{c}} = 0.1$).
        Laser parameters as in \cref{fig:SingleStokes}.
        Pulse durations $\tau_{\tiny\textsf{fwhm}}$: (a,d) 1~ps, (b,e) 2~ps, (c,f) 3~ps.
        \emph{Top row:} Full spectrum of photons with polarisation $\varepsilon_{+}$; 
        \emph{Bottom row:} Cut spectrum (see \cref{fig:NLCSpectra} (right)) of photons with polarisation $\varepsilon_{+}$.
        Labels, colours and line styles as in \cref{fig:GaussianStokes}.
    }
\end{figure}

In \cref{fig:NLCStokes} we show the evolution of the Stokes parameter $\Gamma_2$ for the spectra produced by nonlinear Compton scattering between 10~GeV electrons and a circularly polarised plane-wave laser pulse with dimensionless intensity parameter $\xi_{\tiny\textsf{c}} = 0.1$.
The line styles and colouring match those of the Gaussian case considered previously in \cref{fig:GaussianStokes}.
We note that filtering, besides removing the photons polarised perpendicular to the background ($\varepsilon_{-}$), has two key effects on the energy spectrum.
Firstly, the mean energy of photons emitted close to the electron propagation axis in the $\mathcal{P}_{+}$ distribution increases from $\overline{\omega} \approx 1.5$~GeV (in the unfiltered case) to $\overline{\omega} \approx 1.8$~GeV.
Secondly, the standard deviation decreases from $\sigma \approx 0.38$~GeV to $\sigma \approx 0.14$~GeV, corresponding to 25\% and 8\% of the mean values, respectively.
From our previous discussion of the Gaussian spectrum, we would expect these differences to affect both the final spread of photons and the quality of approximating the averaged spectra by a monoenergetic single photon spectrum with mean photon energy $\omega = \overline{\omega}$.

For the shorter pulses, $\tau_{\tiny\textsf{fwhm}} = 1$~ps, and no filtering, the low-energy tail of the $\mathcal{P}_{+}$ spectrum significantly increases the spread of the $\pm\sigma$ bands as the photons progress through the laser pulse. As photons in the low-energy tail of the spectrum scatter less, an average over the whole (unfiltered) spectrum decreases the Stokes parameter compared to the case with filtering. This also diminishes the separation between the all-orders scattering result (blue) and the LO expansion (orange) in comparison to employing only the filtered spectrum.
In both cases, however, the evolution of the averaged Stokes parameter is well approximated by a single photon of fixed energy $\omega = \overline{\omega}$.
Doubling the duration to $\tau_{\tiny\textsf{fwhm}} = 2$~ps, the larger variance in the unfiltered spectrum begins to have a significant effect, both on the relative spread of the $\pm\sigma$ bands and the quality of the  monoenergetic approximation. Perhaps surprisingly, the Stokes parameter averaged with filtering is still well described by a monoenergetic distribution localised at the mean, and this remains true when the duration is increased to $\tau_{\tiny\textsf{fwhm}} = 3$~ps.

This completes our discussion of the Stokes parameter in the more realistic scenario of initial photons with an energy spread. 
Provided the spectrum is of sufficiently narrow width ($\sigma/\overline{\omega} \sim O(0.1)$), the signature of multiple scattering events survives the averaging of the Stokes parameter over the photon spectrum.
Moreover, in this case the evolution of the distributionally averaged Stokes parameter is well described by assuming a delta energy distribution corresponding to a single photon with energy localised at the spectral mean. 
While this may be unsurprising for Gaussian spectra, which is symmetric around the peak given by the mean $\overline{\omega}$, it is perhaps more surprising for the Compton spectra, which are asymmetric around the mean.

So far, however, we have only considered plane-wave fields. 
In a real experiment, the high-power laser will be focussed to spot sizes on the order of $\mu\text{m}$ which may lead to considerable finite-size and boundary effects.
We therefore need to ensure that the birefringence signal remains detectable upon allowing for a strongly focussed background, and it is to this that we now turn.

\section{Scattering on focussed backgrounds} \label{sec:Focus}

We further generalise our analysis to the physically more realistic scenario of a using a \emph{focussed}  (rather than plane-wave) laser pulse as our phase retarder.
This raises the immediate question of how to calculate the Stokes parameter for the case of non-plane-wave fields.
We know that the optical theorem relates the probability for nonlinear Breit-Wheeler pair creation to the forward scattering part of the polarisation operator, cf.\ \cref{eqn:Ifunc}. Combining a high-energy approximation~\cite{DiPiazza:2016maj} with the LCFA, the probability $\mathcal{P}^{\tiny\tsf{focus}}_{\tiny\tsf{nbw}}$ for nonlinear Breit-Wheeler pair creation in a focussed background field is obtained by \emph{integrating} over the corresponding probabilities, $\mathcal{P}^{\tiny\tsf{pw}}_{\tiny\tsf{nbw}}$, in a plane-wave field, with the latter made local by including transverse structure,
\begin{align}\label{eqn:Focussed}
    \mathcal{P}^{\tiny\tsf{focus}}_{\tiny\tsf{nbw}}
    =
    \int_{-\infty}^{+\infty} \ud^{2} x^{\lcperp}
    \rho_{\gamma}(x^{\lcperp})
    \mathcal{P}^{\tiny\tsf{pw}}_{\tiny\tsf{nbw}}[\chi_{\gamma}(\varphi,x^{\lcperp})]
    \,.
\end{align}
The integral has to be weighted with the transverse areal density, $\rho_{\gamma}(x^{\lcperp})$, of the probe photons.
This amounts to including an integration over $x^{\lcperp}$ and calculating the local value of the photon nonlinearity parameter,
\begin{align}\label{eqn:chiperp}
    \chi_{\gamma}(\varphi,x^{\lcperp})
    =
    \eta
    \xi
    \bigg|f(x^{\lcperp})\frac{\ud g(\varphi)}{\ud \varphi}\bigg|
    \,,
\end{align}
where $f(x^{\lcperp})$ is the transverse beam profile and $g(\varphi)$ is the phase envelope. This suggests to calculate the background induced phase lags as
\begin{align}\label{eqn:PhaseFocus}
    \Delta_{j}
    =
    -
    \frac{(k^{0})^{2}}{m^{2}}
    \frac{1}{\eta}
    \int_{-\infty}^{\varphi} \ud x
    \int_{-\infty}^{+\infty} \ud^{2} x^{\lcperp}
    \rho_{\gamma}(x^{\lcperp})
    \delta n_{j}\big[\chi_{\gamma}(x,x^{\lcperp})\big] 
    \,.
\end{align}
For what follows, we use a focussed Gaussian pulse in the paraxial approximation which corresponds to the gauge potential
\begin{align}\label{eqn:GaugeFocus}
    a_{\mu}(\varphi)
    \to
    a_{\mu}(\varphi,x^{\lcperp})
    =
    \xi
    f(x^{\lcperp})
    g(\varphi)
    \epsilon_{\mu}
    \,,
\end{align}
where $g$ is still the plane-wave pulse \cref{eqn:ggaussian}, modified by the transverse structure
\begin{align}\label{eqn:f}
    f(x^{\lcperp})
    =
    \frac{
        \exp\Big(
            - \frac{(x^{\lcperp})^{2}}{w_{0}^{2} (1 + \zeta^{2})}
        \Big)
    }{\sqrt{1 + \zeta^{2}}}
    \,,
\end{align}
with beam radius $w_{0}$, Rayleigh length $z_{r} = \pi w_{0}^{2} / \lambda$ and dimensionless coordinate $\zeta = z/z_{\tiny\tsf{r}}$.

In a real laser system the key parameters are the laser energy, $E_{\tiny{\tsf{laser}}}$, and the central wavelength, $\lambda$.
Using different optics and setups, parameters such as the pulse length and peak intensity can (in principle) be arbitrarily adjusted. To distinguish multiple scattering events we would like to use a laser with a given energy and increase the pulse duration, hence the interaction time, $\tau$. This will in turn decrease the peak power, $P = E_{\tiny{\tsf{laser}}}/\tau$ of the laser pulse. For a physically sensible comparison of different scenarios, we will keep the peak intensity, $I_0$, or dimensionless intensity parameter $\xi_{0}$, fixed while varying the pulse duration. This can be done by suitably adjusting the beam waist, $w_{0}$, as $I_0 \sim P/ w_0^2$.
Let us consider parameters specific to the L4-ATON laser system at ELI Beamlines with $E_{\tiny{\tsf{laser}}} = 1.5$~kJ and $\lambda = 1.05$~$\mu$m. Thus, each cycle of the pulse has a duration $\tau_{\text{cycle}} = \lambda/c \sim 3.5$~fs. The laser system will have a typical operational mode working at 10~PW peak power output with $150$~fs pulse duration, corresponding to $N \sim 43$ cycles. Adjusting the optics should allow the pulse duration to reach the ps regime. A pulse duration of $\tau \sim 10$~ps, for example, would correspond to $N \sim 3000$ cycles. The intensity of such a long and defocussed pulse will be reduced, though.

Following~\cite{DiPiazza:2016maj}, recovering the plane-wave result from \cref{eqn:PhaseFocus} involves taking $\rho_{\gamma}(x^{\lcperp}) = \rho_{\gamma}$ constant and $w_{0} \to \infty$ in \cref{eqn:f}.
The transverse integration region, $\Sigma$, is chosen such as to ensure that the peak power in both the focussed and plane-wave cases are equal, $\Sigma \approx \pi w_{0}^{2}/2$. For the plane-wave case, we employ the transverse areal probe density $\rho_{\gamma} = \Sigma^{-1}$, while for the focussed case we use $\rho_{\gamma}(x^{\lcperp}) = \Sigma^{-1} \theta(w_{0} - |x^{\lcperp}|)$. We thus assume that the photons are confined to a uniform disk with the same area as the focal spot. Furthermore, we neglect the relative transverse momentum between photons in the bunch. These approximations are valid as long as the propagation distance, $d$, from the photon source to the laser focus is large compared to the beam waist, $d \gg w_{0}$. This should be well-satisfied as typical distances will be of order $d \sim O(\text{cm}-\text{m})$, much larger than the  focal spot radius, $w_{0} \sim O(\mu\text{m})$.

\begin{figure}[t!!]
    \centering
    \includegraphics[width=0.99\textwidth,trim={0.0cm 0.0cm 0.0cm 0.0cm},clip=true]{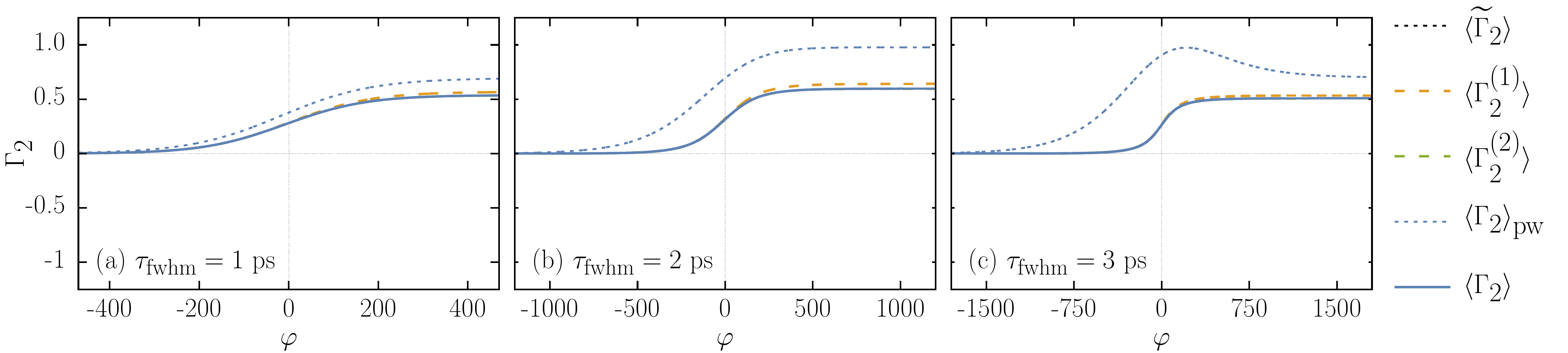}
 
    \caption{\label{fig:FocusStokes} 
        Evolution of the Stokes parameter, $\Gamma_2$, with the laser phase, $\varphi$, for probe photons produced by Compton scattering (between 10~GeV electrons and a circularly polarised laser pulse with $\xi_{\tiny\tsf{c}} = 0.1$). \emph{Laser}: linearly polarised with Gaussian focus and peak dimensionless intensity parameter $\xi_{0} = 50$, wavelength $\lambda = 1.05$~$\mu\text{m}$
        and three different choices for full-width-half-maximum pulse duration and waist radius, $\tau_{\tiny\tsf{fwhm}}$: (a) 1~ps and 3.92 $\mu$m, (b) 2~ps and 2.77 $\mu$m, (c) 3~ps and 2.26 $\mu$m.
        \emph{Probe photon spectrum}: filtered as in \cref{fig:NLCSpectra} and \cref{sec:EnergySpectraNLC}.
        \emph{Solid lines}: All-orders Stokes parameter with pair creation.
        \emph{Dot-dashed lines}: Ditto without pair creation.
        \emph{Dashed lines}: $(2n+1)$-fold scattering results for $n = 1$ (orange), $n = 2$ (green), and $n = 3$ (red). 
        \emph{Dotted lines}: all-orders Stokes parameter in a plane-wave field.
    }
\end{figure}

In \cref{fig:FocusStokes} we show the evolution of the distributionally averaged Stokes parameter for photons interacting with a Gaussian focussed pulse, \cref{eqn:GaugeFocus}.
The photons are, as before, produced by nonlinear Compton scattering of monoenergetic 10~GeV electrons with a weak ($\xi_{\tiny\tsf{c}} = 0.1$) circularly polarised laser pulse, and filtered with the combination of a transverse momentum cut at $r_{\tiny\tsf{cut}}^{2} = (8/25) \eta_{e} s_{\star,1}$ and low-energy filter, cf.\ \cref{fig:NLCSpectra} and \cref{sec:EnergySpectraNLC}.
The secondary laser, which acts as a linear phase retarder on the propagating photons, has a pulse energy of $E_{\tiny{\tsf{laser}}} = 1.5$~kJ and wavelength $\lambda = 1.05~\mu$m, corresponding to the experimental parameters available with the upcoming ELI Beamlines L4-ATON laser system.
We consider the peak dimensionless intensity parameter of the focussed laser pulse to be $\xi_{0} = 50$ and again increase the pulse duration, $\tau_{\tiny\tsf{fwhm}} = (1,2,3)$~ps.
Maintaining an equivalent peak intensity whilst holding the pulse energy constant requires focussing from, respectively, $w_{0} = (3.9,2.8,2.2)~\mu$m as the pulse duration is increased.
An immediate consequence of this is the shortening of the Rayleigh length, since $z_{\tiny\tsf{r}} \propto w_{0}^{2}$, which results in a decrease in the longitudinal interaction length. (It has been shown in the literature \cite{King:2012aw,King:2018wtn} that the relevant longitudinal length scale is $\trm{min}\{\tau,z_{r}\}$, and here, $z_{r} \ll \tau$.) In this regime, the interaction volume, $\Omega \approx z_r w_0^2 \pi  = \pi^2 w_0^4/\lambda \ll \tau w_0^2 \pi $.
This suppresses the multiple scattering signal calculated in the preceding section for the case of plane-wave fields, for the parameters we have considered. 
There are two possible routes out of this.
The first is to increase the photon energy, and decrease $\xi_{0}$, such that the maximum $\chi$ remains the same but the Rayleigh length is sufficiently long to allow for a longer interaction time with the propagating photons.
The second possibility is to instead consider alternative set-ups with a longer effective interaction time, such as achieved if e.g. multiple pulses or special configurations such as flying focus beams \cite{Longhi:04,Ramsey.Phys.Rev.A.2023} are employed.

\section{Discussion} \label{sec:Discussion}

In this work we have outlined details of a possible experimental scenario that could be suitable for measuring vacuum birefringence, by combining photons from a polarised Compton source with a high-energy laser pulse stretched to picosecond durations.
Such a scheme will be possible using the unique combination of the ELI Beamlines L3-HAPLS and L4-ATON laser systems.
We have suggested an approach for discerning $(2n+1)$-fold scattering (for different $n$) of photons in the laser pulse, and demonstrated how various experimental conditions, such as the photon energy distribution and laser focussing, can influence the measurable signal.

We demonstrated that when focussing is taken into account, the multi-scattering signals which were seen in the plane-wave approximation diminish. 
Ideally, to observe multiple scattering in an experiment, one needs to have a long interaction time where the laser energy is being most effectively utilised.
For the single pulse considered here, the tight focussing in the long duration case causes the pulse to quickly diverge from the focus, and only a small portion of the total pulse energy is effectively `seen' by the propagating photons.
One way to increase the interaction time is to split the total available laser energy into $n$ identical pulses of shorter duration and then stagger the position of the foci for each pulse by a distance on the order of the Rayleigh length, $z_{r}$, such that the effective interaction length becomes on the order of $L \sim 2 n z_{\tiny\tsf{r}}$. 

Another way to increase the interaction time would be exploring the use of more exotic focussing schemes, in particular the so-called flying focus~\cite{Ramsey.Phys.Rev.A.2023}. This is an exact solution to Maxwell's equations which describes a laser pulse with focus moving at an arbitrary, controllable, speed.
Through matching the speed of the drifting focus with that of the incident particles, one can essentially ensure that the particles always remain at the peak field strength. This has been shown to lead to an enhancement of, for example, radiation reaction effects by electrons~\cite{Formanek:2021bpw}.
We suggest that a flying focus would also be useful for birefringence experiments, as increasing the time the probe photons are exposed to the peak field should lead to stronger signals.

There are a number of other avenues which could be explored in future work, in particular to aid the further development of an experiment based on our ideas.
An immediate extension is using a non-monoenergetic electron beam to source our polarised photons.
Laser wakefield acceleration has demonstrated the ability to produce quasi-monoenergetic electron beams of multi-GeV energies with energy spreads of 6-10\%~\cite{Wang:2013zpo,Leemans:2014lka,Gonsalves:2019wnc}. 
These results, however, are rather sensitive to the experimental set-up, with the inclusion of multiple acceleration stages and higher density targets resulting in broader energy spectra with spreads on the order of 30-50\%~\cite{Kim:2013zsa,Cole:2017zca,Poder:2017dpw}.
Furthermore, as well as having a spread in the energy distribution, laser wakefield accelerated electrons can also have relatively large beam divergences in comparison to conventionally accelerated electrons~\cite{Kim:2021jjy}.
To ensure the accuracy of the modelling of the photon production stage, these factors should be included.

The current work highlights that strong-field QED effects can also be explored with high \emph{energy} laser pulses, rather than just considering the pulse \emph{intensity} as the only relevant parameter.
The ensuing increase in pulse duration should also be relevant for other processes such as higher-order Compton scattering or trident pair production.

\section*{Acknowledgements}

BK acknowledges the hospitality of the DESY theory group and support from the Deutsche Forschungsgemeinschaft (DFG, German Research Foundation) under Germany’s Excellence Strategy – EXC 2121 “Quantum Universe” – 390833306.
AJM and SVB are supported by the project ``Advanced Research Using High Intensity Laser Produced Photons and Particles'' (ADONIS) CZ.02.1.01/0.0/0.0/16\_019/0000789 from European Regional Development Fund (ERDF).
AJM thanks Tae Moon Jeong for useful discussions.

\appendix

\section{Stokes parameters \label{app:Stokes}}

Here we provide some more detail on the Stokes parameters used to probe the polarisation asymmetry in the phase accrued by a probe photon traversing a linearly polarised background. The Jones vector, \cref{eqn:FieldPolarisation}, describes the phase dependent polarisation of a photon propagating through a strong linearly polarised plane-wave field, defined in terms of two complex coefficients, $c_{x}(\varphi)$ and $c_{y}(\varphi)$, which give the components of the polarisation along the directions specified by the eigenstates of the background.
These components can be mapped onto a $2 \times 2$ polarisation density matrix $\mathcal{M}$ with components $\mathcal{M}_{ij} = c_{i}^{*} c_{j}, ~ i,j \in \{x,y\}$, which combines them into four real parameters, $S_{0},S_{1},S_{2},S_{3}$, known as the Stokes parameters\footnote{The Stokes parameters are sometimes denoted $(S_{0},S_{1},S_{2},S_{3}) \equiv (I,Q,U,V)$.}: 
\begin{align}\label{eqn:StokesParams}
    S_{0} 
    =
    &
    |c_{x}|^{2} + |c_{y}|^{2}
    \,,
    \quad
    &
    S_{1} 
    =
    &
    |c_{x}|^{2} - |c_{y}|^{2}
    \,,
    \quad
    &
    S_{2} 
    =
    &
    2 \RE [c_{x}^{*} c_{y}]
    \,,
    \quad
    &
    S_{3} 
    =
    &
    2 \IM [c_{x}^{*} c_{y}]
    \,,
\end{align}
such that,
\begin{align}\label{eqn:StokesMatrix}
    \mathcal{M}
    = S_0 \mathbb{I} + S_i \sigma_i =
    \frac{1}{2}
    \begin{pmatrix}
        S_{0} + S_{1} ~& S_{2} - i S_{3} \\
        S_{2} + i S_{3} ~& S_{0} - S_{1}
    \end{pmatrix}
    \,.
\end{align}
For fully polarised light, the polarisation matrix $\mathcal{M}$ has vanishing determinant, $\det \mathcal{M} = S_0^2 - \mathbf{S}^2 =0$, and thus can be identified with a null Stokes 4-vector $S^\mu=(S_{0},\mathbf{S})$, with $S^2 = 0$. In the general case of partial polarisation, the polarisation matrix can be decomposed into completely polarised and unpolarised components\cite{Roemer:2005}, $\mathcal{M} = \mathcal{M}_p + \mathcal{M}_u$.

Physically, $S_{0}$ gives the photon survival probability in the background field, such that in the absence of pair creation $S_{0} \to 1$, while the remaining Stokes parameters give the degree of polarisation along different polarisation axes.
The Stokes parameter $S_{1}$ represents the asymmetry of the photon polarisation in the $\ket{x,y}$ basis, since the probability that a photon will be measured in the states $\ket{x}, \ket{y}$ is just $\prob_{j} = |c_{j}|^{2}$. As explained in the main text, $\delta\phi_{j}(\varphi)$ is in general complex valued, so expanding it into real and imaginary parts, $\delta\phi_{j} = \RE \delta\phi_{j} + i \, \IM \delta\phi_{j}$, and suppressing the explicit dependence on $\varphi$ for brevity, the probabilities are $\prob_{x} = \mbox{e}^{- 2 \IM(\delta\phi_{x})} \cos^{2} \theta_{0}$ and $\prob_{y} = \mbox{e}^{- 2 \IM(\delta\phi_{y})} \sin^{2} \theta_{0}$.
The dependence on the real part of the phase lags $\delta\phi_j$ has vanished, and in the absence of pair creation (i.e. if $\IM (\delta\phi_{j}) = 0$) the probabilities do not evolve in $\varphi$ and are completely determined by the initial angle $\theta_0$. The same is true for the Stokes parameter $S_1$,  defined by

\begin{align} \label{S.1a}
  S_1 = \prob_{x} - \prob_{y} 
  = 
  \mbox{e}^{- 2 \IM\delta\phi_{x}} \cos^{2} \theta_{0}
  -
  \mbox{e}^{- 2 \IM\delta\phi_{y}} \sin^{2} \theta_{0}
  \;,
\end{align}
which in the absence of pair production reduces to,
\begin{align} \label{S.1}
  \lim\limits_{\IM(\delta \phi_x),\IM(\delta \phi_y) \to 0} S_1 = \cos 2 \theta_0 \; .
\end{align}
In particular, if the photon was initially polarised in an eigenstate of the background ($\theta_{0} = 0$ or $\pi/2$), then it \emph{remains} in that polarisation eigenstate and $S_1 = 1$.

Let us note that the inclusion of the imaginary part of the polarisation operator leads to an exponential suppression of the probability when $\IM (\delta\phi_{j})$ becomes of the order of 
unity. Of course, this is nothing but absorption: photons are being removed from the initial state via  pair production. These pairs would subsequently be accelerated and radiate further photons (final-state radiation) at the same perturbative order in the fine-structure constant $\alpha$ as the dispersive process we are considering. As we are only interested in the latter, we need to keep pair production and the ensuing secondary radiation at a negligible rate. This is in line with the refractive index approach, which is intrinsically based on the background field method, and cannot account for any secondary radiation. We thus must limit ourselves to parameter regimes where pair production is not significant.

According to (\ref{S.1a}), the Stokes parameter $S_{1}$ is blind to (dispersive) vacuum polarisation effects encoded in the real parts of the phase lags, $\delta\phi_j$. We must therefore consider a measurement of the other Stokes parameters. For the Stokes parameter $S_2$ we introduce the linear superposition basis $\ket{1,2}$ where $\ket{1} = (\ket{x} + \ket{y})/\sqrt{2}$ and $\ket{2} = (\ket{x} - \ket{y})/\sqrt{2}$. Projecting the polarisation state \cref{eqn:FieldPolarisation} onto this basis, the Stokes parameter $S_{2}$ measures the asymmetry $S_{2} = \prob_{1} - \prob_{2}$, in analogy with (\ref{S.1a}). Explicitly, these probabilities are
\begin{align}\label{eqn:Prob12}
    \prob_{1}
    =
    &
    \frac{\mbox{e}^{- \IM(\delta\phi_{x} + \delta\phi_{y})}}{2}
    \Big[
        \mbox{e}^{- \IM(\Delta)}
        \cos^{2}\theta_{0}
        +
        \sin(2\theta_{0})
        \cos\left[\RE(\Delta) - \psi_{0}\right]
        +
        \mbox{e}^{+ \IM(\Delta)}
        \sin^{2}\theta_{0}
    \Big]
    \,,
    \\
    \prob_{2}
    =
    &
    \frac{\mbox{e}^{- \IM(\delta\phi_{x} + \delta\phi_{y})}}{2}
    \Big[
        \mbox{e}^{- \IM(\Delta)}
        \cos^{2}\theta_{0}
        -
        \sin(2\theta_{0})
        \cos\left[\RE(\Delta) - \psi_{0}\right]
        +
        \mbox{e}^{+ \IM(\Delta)}
        \sin^{2}\theta_{0}
    \Big]
    \,,
\end{align}
where we have used the \emph{difference in phase difference} defined in (\ref{eqn:Diff}), $\Delta(\varphi) \equiv \delta\phi_{x}(\varphi) - \delta\phi_{y}(\varphi) \,.$

The difference of the probabilities gives the second Stokes parameter, 
\begin{align}\label{S.2}
    S_{2}
    =
    &
    \mbox{e}^{- \IM(\delta\phi_{x} + \delta\phi_{y})}
    \sin(2\theta_{0})
    \cos\left[\RE(\Delta) - \psi_{0}\right]
    \,,
\end{align}
to be compared with $(\ref{S.1a})$.

We again find the result that an incoming eigenstate of the background (when $\theta_{0} = 0$ or $\pi/2$) remains an eigenstate, this time characterised by $S_{2} = 0$. In this case, the polarisation remains constant, and the transition probabilities from the $\ket{x,y}$ basis to the $\ket{1,2}$ basis are $\prob_{1} = \prob_{2} = \mbox{e}^{-2 \IM(\phi_{j})}/2$, where $j=x$ for $\theta_{0} = 0$, and $j=y$ for $\theta_0 = \pi/2$.

For general initial data, ($\theta_{0}$, $\psi_{0})$, however, the Stokes parameter (\ref{S.2}) explicitly depends on the real part of the polarisation operator through $\Delta$. Hence, as the photons propagate through the background and evolve with $\varphi$, vacuum birefringence generates a phase lag $\RE\Delta(\varphi)$ between the two linear polarisation components, which causes the Stokes parameter to oscillate in $\varphi$. The magnitude of these oscillations is gradually damped by the decay of photons into electron-positron pairs through the nonlinear Breit-Wheeler process, as encapsulated in the decay factor $\mbox{e}^{-\IM(\delta\phi_{x} + \delta\phi_{y})}$ (although we reiterate that we have limited this decay to be negligible in the present work).

For the Stokes parameter $S_{3}$ we measure the helicity of the photon by projecting onto helicity eigenstates $\ket{\pm} = \ket{x} \pm i \ket{y}$. The ensuing asymmetry is now $S_{3} = \prob_{+} - \prob_{-}$, with the probabilities given by
\begin{align}\label{eqn:Prob+}
    \prob_{+}
    =
    &
    \frac{\mbox{e}^{- \IM(\delta\phi_{x} + \delta\phi_{y})}}{2}
    \Big[
        \mbox{e}^{- \IM(\Delta)}
        \cos^{2}\theta_{0}
        -
        \sin(2\theta_{0})
        \sin(\RE(\Delta) - \psi_{0})
        +
        \mbox{e}^{+ \IM(\Delta)}
        \sin^{2}\theta_{0}
    \Big]
    \,,
    \\\label{eqn:Prob-}
    \prob_{-}
    =
    &
    \frac{\mbox{e}^{- \IM(\delta\phi_{x} + \delta\phi_{y})}}{2}
    \Big[
        \mbox{e}^{- \IM(\Delta)}
        \cos^{2}\theta_{0}
        +
        \sin(2\theta_{0})
        \sin(\RE(\Delta) - \psi_{0})
        +
        \mbox{e}^{+ \IM(\Delta)}
        \sin^{2}\theta_{0}
    \Big]
    \,.
\end{align}
The difference of the probabilities\footnote{We note that setting $\theta_{0} = \pi/4$, $\psi_{0} = 0$ and $\IM\phi_{j} = 0$ in \cref{eqn:Prob+,eqn:Prob-} we recover the results in King and Elkina\cite{King:2016jnl}.} is the asymmetry
\begin{align}\label{eqn:S3}
    S_{3}
    =
    &
    -
    \mbox{e}^{- \IM(\delta\phi_{x} + \delta\phi_{y})}
    \sin(2\theta_{0})
    \sin\left[\RE(\Delta) - \psi_{0}\right]
    \,,
\end{align}
to be compared with (\ref{S.1a}) and (\ref{S.2}). For completeness, we list the Stokes parameter $S_{0}$ representing total photon number or intensity,
\begin{align}\label{eqn:S0}
    S_{0}
    =
    &
    \mbox{e}^{- \IM(\delta\phi_{x} + \delta\phi_{y})}
    \Big[
        \mbox{e}^{- \IM(\Delta)}
        \cos^{2}(\theta_{0})
        +
        \mbox{e}^{+ \IM(\Delta)}
        \sin^{2}(\theta_{0})
    \Big] 
    \,.
\end{align}
This must be the sum of the probabilities in any chosen orthonormal measurement basis, i.e., $S_{0} = \prob_{x} + \prob_{y} = \prob_{1} + \prob_{2} = \prob_{+} + \prob_{-}$.
As a measure of the total photon number, or intensity, $S_{0}$ is always independent of $\RE(\Delta)$, since birefringence cannot influence the total number of photons. However, $S_0$ does depend on $\IM(\Delta)$, since pair creation implies the removal of photons from the initial state.

It will be useful to define the normalised Stokes parameters $\Gamma_{j} = S_{j}/S_{0}, ~ j \in \{1,2,3\}$,
\begin{align} \label{eqn:NormalStokes}
    \Gamma_{1}
    = 
    &
    \frac{
        \mbox{e}^{- \IM(\Delta)} \cos^{2} \theta_{0} - \mbox{e}^{+ \IM(\Delta)} \sin^{2} \theta_{0} 
    }{\mbox{e}^{- \IM(\Delta)} \cos^{2}\theta_{0} + \mbox{e}^{+ \IM(\Delta)} \sin^{2}\theta_{0}}
    \;,
    \\
    \Gamma_{2}
    =
    &
    \frac{
        \sin(2\theta_{0}) \cos\left[\RE(\Delta) - \psi_{0}\right]
    }{\mbox{e}^{- \IM(\Delta)} \cos^{2}\theta_{0} + \mbox{e}^{+ \IM(\Delta)} \sin^{2}\theta_{0}}
    \,,
    \\
    \Gamma_{3}
    =
    &
    -
    \frac{
        \sin(2\theta_{0}) \sin\left[\RE(\Delta) - \psi_{0}\right]
    }{\mbox{e}^{- \IM(\Delta)} \cos^{2}\theta_{0} + \mbox{e}^{+ \IM(\Delta)} \sin^{2}\theta_{0}}
    \,.
\end{align}
One can see that the Stokes parameters $\Gamma_{2,3}$ will be maximised in the absence of pair production when $\theta_{0} = \pm \pi/4$, which corresponds to the photon initially being polarised in $\ket{1,2}$ or $\ket{+,-}$.
Furthermore, if we expand the normalised Stokes parameters $\Gamma_{2,3}$ for small $\IM(\Delta)$ (we define $\widetilde{\Gamma}_{2,3}$ to correspond to the numerators of $\Gamma_{2,3}$ which are independent of the imaginary part of $\Delta$),
\begin{align} \label{eqn:NormalStokesExpand}
    \Gamma_{2,3}
    =
    &
    \widetilde{\Gamma}_{2,3}
    \bigg[
        1
        +
        \IM(\Delta)
        \cos 2\theta_{0}
        +
        \frac{1}{2}
        \left[\IM(\Delta)\right]^{2}
        \cos 4\theta_{0}
        +
        \dots
    \bigg]
    \,,
\end{align}
we see that the choice of $\theta_{0} = \pm \pi/4$ to maximise elastic scattering causes pair creation to enter only at order $\left[\IM(\Delta)\right]^{2}$. 
Furthermore, when pair-creation in the regime $\xi \gg 1$, $\chi \ll 1$ is less prevalent than elastic scattering (the regime in which the locally constant field approximation is valid), it is exponentially suppressed in \cref{eqn:NormalStokesExpand}.
Therefore, in the current study, by choosing the angle $\theta_{0}$ to maximise the elastic scattering signal, pair creation can only represent a small correction.


\providecommand{\noopsort}[1]{}

\end{document}